\documentclass[
twocolumn,
hf
]{ceurart}
\usepackage{inputenc}
\usepackage{tabularx}
\usepackage{listings}
\usepackage{subcaption}
\begin{document}

\copyrightyear{2025}

\copyrightclause{Copyright for this paper by its authors. Use permitted under Creative Commons License Attribution 4.0 International (CC BY 4.0).}

\conference{RecSys in HR'25: The 5th Workshop on Recommender Systems for Human Resources, in conjunction with the 19th ACM Conference on Recommender Systems, September 22--26, 2025, Prague, Czech Republic.}

\title{Explained, yet misunderstood: How AI Literacy shapes HR Managers’ interpretation of User Interfaces in Recruiting Recommender Systems}

\author[1]{Yannick Kalff}[%
orcid=0000-0003-1595-175X,
email=yannick.kalff@htw-berlin.de,
url=https://yannickkalff.de,
]
\cormark[1]
\fnmark[1]
\address[1]{HTW Berlin University of Applied Sciences,
  Treskowallee 8, 10318 Berlin, Germany}

\author[1]{Katharina Simbeck}[%
orcid=0000-0001-6792-461X,
email=katharina.simbeck@htw-berlin.de,
url=https://iug.htw-berlin.de,
]
\fnmark[1]

\cortext[1]{Corresponding author.}
\fntext[1]{These authors contributed equally.}

\begin{abstract}
	AI-based recommender systems increasingly influence recruitment decisions. Thus, transparency and responsible adoption in Human Resource Management (HRM) are critical. This study examines how HR managers’ AI literacy influences their subjective perception and objective understanding of explainable AI (XAI) elements in recruiting recommender dashboards. In an online experiment, 410 German-based HR managers compared baseline dashboards to versions enriched with three XAI styles: important features, counterfactuals, and model criteria. Our results show that the dashboards used in practice do not explain AI results and even keep AI elements opaque. However, while adding XAI features improves subjective perceptions of helpfulness and trust among users with moderate or high AI literacy, it does not increase their objective understanding. It may even reduce accurate understanding, especially with complex explanations. Only overlays of important features significantly aided the interpretations of high-literacy users. Our findings highlight that the benefits of XAI in recruitment depend on users’ AI literacy, emphasizing the need for tailored explanation strategies and targeted literacy training in HRM to ensure fair, transparent, and effective adoption of AI.
\end{abstract}


\maketitle

\section{Introduction}
Artificial intelligence (AI)-based recommender systems have become widespread in recruitment \cite{Zhai.2024, Malik.2023}. Recommender systems are software applications that use artificial intelligence techniques to analyze data and provide specific suggestions or predictions to users. AI-based systems typically assist in discovering promising talents for development, identifying the most suitable candidates for a job opening, or assigning the right employees to projects based on their skill sets. In human resource management (HRM), these tools promise to accelerate processes, reduce human bias, and ground decisions in objective data \cite{Drage.2022, Kelan.2024}. Current trends, such as HR Analytics and People Analytics, integrate AI to offer a broader promise of analytical rigor, predictive opportunities, and prescriptive recommendations for informed decision-making and actions \cite{Edwards.2024}, alongside technological modes of control \cite{Klopper.2023}. In recruiting, AI recommender systems directly influence decisions about individuals, making them the subject of regulations, such as the EU AI Act \cite{EuropeanCommission.2021} and the GDPR \cite{EuropeanParliament.2016}. Moreover, HR systems have faced severe criticism for fairness issues and biased recommendations \cite{Du.2024, Fabris.2024, Simbeck.2019, Kochling.2021}.

To mitigate potential biases and, equally important, to make optimal decisions, HR managers must understand the underlying data models and the mechanisms by which individual recommendations are generated. Explainable AI (XAI) techniques aim to make “black-box” models interpretable when they are opaque---due to complexity or proprietary constraints \cite{Molnar.2022, Bhatt.2020}. 

XAI methods offer interpretable, context-specific explanations for model decisions \cite{Speith.2022, Khalili.2023}. In recruitment, these explanations can clarify why a candidate’s application is ranked highly or why specific competencies are flagged during the CV parsing process. This transparency is essential for HR managers who often have non‐technical backgrounds and are responsible for legally and ethically sound decisions that comply with anti‐discrimination laws. At the same time, from a human resources management perspective, their decisions must be economically sensible and strategically appropriate for the company. A lack of transparency in AI elements or data, combined with unrecognized distortions, can lead users to incorrect conclusions. The issue is amplified by providers and developers, as transparency and explanations of AI interfaces remain the exception in practice. Lacking transparency often seems to be a deliberate UI design decision (three exemplary dashboards can be found in the appendix Figure~\ref{fig:dashboards}--\ref{fig:dashboards3}).

However, attaching explanation widgets to a recruitment dashboard does not guarantee impact. For XAI to be effective, HR managers must decode and critically evaluate the provided information \cite{Bauer.2023}. We contend that AI literacy---a combination of knowledge, skills, and attitudes that enables individuals to understand and assess AI systems \cite{Laupichler.2023, Carolus.2023}---directly affects both the subjective and objective effectiveness of XAI. Subjectively, AI literacy influences how helpful, trustworthy, and accessible explanations appear. Objectively, it affects accurate factual understanding that HR managers present when they interpret and act on the information provided by AI dashboards.

To investigate this effect, we conducted an experiment with  410 German-based HR managers, who compared a baseline AI dashboard with versions enriched by three explanation styles: important features (a simplified feature importance approach), counterfactual explanations, and global model criteria summaries \cite{Guidotti.2019, Bodria.2023}. Drawing on a genuine recruitment ranking tool that exists on the market, we measured participants’ perceived trust, usability, and assessment quality for each existing dashboard variant. Further, the participants assessed statements on the dashboards with correct or false answer possibilities. We address two guiding research questions:
\begin{itemize}
	\item[\textbf{RQ1}] How do HR managers' subjective perceptions of a recruiting recommender system change when adding explainable AI elements, and does this effect differ across different levels of AI literacy?
	\item[\textbf{RQ2}] How does HR managers' objective understanding of a recruiting recommender system change when adding explainable AI elements, and does this effect differ across different levels of AI literacy?
\end{itemize}
Our results show that higher AI literacy is associated with greater perceived usefulness and transparency of XAI‐enhanced dashboards. Paradoxically, higher AI literacy also corresponds to a lower objective understanding of the user interfaces. These findings suggest that the benefits of XAI depend critically on users’ AI literacy levels, though self-assessed AI literacy may be prone to overconfidence.

The article is structured as follows: First, we review the literature on AI literacy and XAI, with a focus on their intersection in recruitment contexts (\ref{sec:research}). Next, we present our methodological approach (\ref{sec:design}) and empirical findings (\ref{sec:findings}). We discuss the theoretical and practical implications for responsibly embedding explainable recommender systems in HRM that arise from AI literacy and its impact on subjective perception and objective understanding (\ref{sec:discussion}). We conclude with an outlook on future research topics that can be derived from our findings (\ref{sec:conclusion}).

\section{State of Research}\label{sec:research}
The scholarly discourses on AI literacy and explainable AI (XAI) have so far evolved independently (with few exceptions \cite{Bhat.2024}). AI literacy research emphasizes the competencies required to comprehend, evaluate, and interact with AI‐enabled tools \cite{Lintner.2024}. Several authors have developed scales to assess individuals’ abilities to recognize, understand, apply, and critically or ethically evaluate AI systems \cite{Laupichler.2023, Carolus.2023, Ng.2024}. Empirical studies underscore the need for contextualized training that embeds domain-relevant examples and ethical deliberation, arguing that generic digital skills programs fall short of preparing professionals for AI-mediated work \cite{Pinski.2024, Bassellier.2003}. In HRM, such contextualization is particularly vital, as recruitment decisions carry significant strategic, legal, or ethical weight with impact on companies' success, diversity, equity, and the organization's reputation. A high level of AI literacy would include a foundational understanding of the technical principles underlying AI systems---such as training procedures, or the critical role of data quality---and the ability to use AI tools effectively in appropriate contexts. Proficiency in AI literacy would further indicate awareness of AI's limitations and boundaries, including ethical considerations and potential grey areas, and the necessity of human oversight. Moreover, high levels of AI literacy involve the ability to recognize AI-driven processes, critically assess the outputs generated by such systems, and accurately identify their capabilities and limitations.

XAI research, by contrast, focuses on designing algorithms, systems, and user interfaces that render AI decisions and recommendations transparent and understandable \cite{Clinciu.2019, Gunning.2019}. This approach treats users as accountable agents who must comprehend, evaluate, and, if necessary, correct AI outputs \cite{Chowdhury.2023a}. Initially, XAI addressed the needs of ML engineers and developers seeking to understand and debug complex AI models \cite{Bhatt.2020}. Recent developments in XAI have extended the audience for explanations and established that explanations need to account for users’ roles, backgrounds, and prior technical knowledge \cite{Cambria.2023}. Researchers distinguish between global explanations, which clarify an entire model’s logic, and local explanations, which justify individual decisions \cite{Guidotti.2019}. Adequate XAI should also draw on domain-specific knowledge---for example, highlighting key CV attributes or motivational letter elements that influenced an AI recommendation \cite{Holzinger.2022b}.

Although contextualization and audience adaptation have been emphasized, little attention has been paid to how users' AI literacy affects the efficacy of XAI elements. Explanations are meaningful only if recipients possess the cognitive and critical frameworks to interpret them: Global explanations of feature weights presuppose familiarity with model training and evaluation metrics, whereas local explanations of ranking positions require understanding how feature contributions differ across cases. Without critical literacy, HR managers may overlook XAI elements or succumb to confirmation bias,  disregarding explanations that challenge their prior assumptions. Similarly, deficient practical AI literacy can lead users to fail to recognize when they are interacting with AI, thereby undermining the necessary critical scrutiny. Consequently, even well‐designed explanations may fail to foster appropriate trust or may inadvertently reinforce erroneous mental models.

A significant research gap is the lack of systematic insight into how non-technical experts, such as HR managers, perceive XAI subjectively (for example, in terms of perceived usefulness or trustworthiness), how XAI contributes to their objective understanding (such as the accurate interpretation of AI outputs), and how the effectiveness of XAI varies according to different levels of AI literacy among HR managers. Addressing this gap is crucial for three reasons. First, without insight into user comprehension, organizations risk deploying XAI that engenders misplaced trust or unwarranted skepticism. Second, regulatory frameworks increasingly mandate transparency, but compliance depends on decision-makers' ability to understand the provided explanations \cite{Kuhl.2023a}. Third, investments in AI systems for HR---especially recruiting recommender systems---must not only incorporate explainability features but also ensure that HR professionals receive the AI literacy training necessary to operate these tools responsibly and sustainably.

\section{Research design}\label{sec:design}
In an experiment, we queried 427 HR managers in Germany. After excluding implausible cases, the study retained 410 valid responses. We assessed the HR managers’ AI literacy using the \enquote{scale for the assessment of non-experts’ AI literacy} (SNAIL) \cite{Laupichler.2023}. The scale comprises three dimensions---\textit{technical understanding} (TU), \textit{critical appraisal} (CA), and \textit{practical application} (PA)---each measured with ten Likert-scaled items from which we picked five. We selected the items from the full 30-item SNAIL scale based on their relevance to the HR domain (cf. Table~\ref{tab:items} for an overview of items and individual statistics).

The scale demonstrated high reliability, with strong Cronbach’s $\alpha$ values across all three dimensions (Table~\ref{tab:ailit}).  The dimensions exhibited strong collinearity, indicating that participants who scored low/high on one dimension tended to do so on the other dimensions as well. For further analyses, we classified participants---low, medium, and high AI literacy---by dividing the scale into three equal intervals (Table~\ref{tab:ailitgroups}).

We conducted an experiment to assess the effect of XAI elements on users. We researched interfaces and dashboards of existing HR tool vendors that advertise AI functions (for example, Figure~\ref{fig:dashboards}-\ref{fig:dashboards3}). Those vendors present their dashboards as advertisements and use cases, usually on the companies' websites. Figure~\ref{fig:dashboards} shows an application recommender system with several active applications, their grading, skill match, and additional personal information. On closer examination, the criteria for ranking grade and skill match, and consequently, the resulting recommendation, seem ambiguous. For example, it is unclear why the recommended first-place candidate receives an A grade, despite having a substantially lower skill match than subsequent candidates. There is no explanation of how the sorting was conducted or why the ranking, which initially appears implausible, could nevertheless be justified. Moreover, it is uncertain whether the results reflect possible errors in the underlying AI system.
\begin{figure*}
	\begin{subfigure}{0.45\textwidth}
		\includegraphics[width=\textwidth]{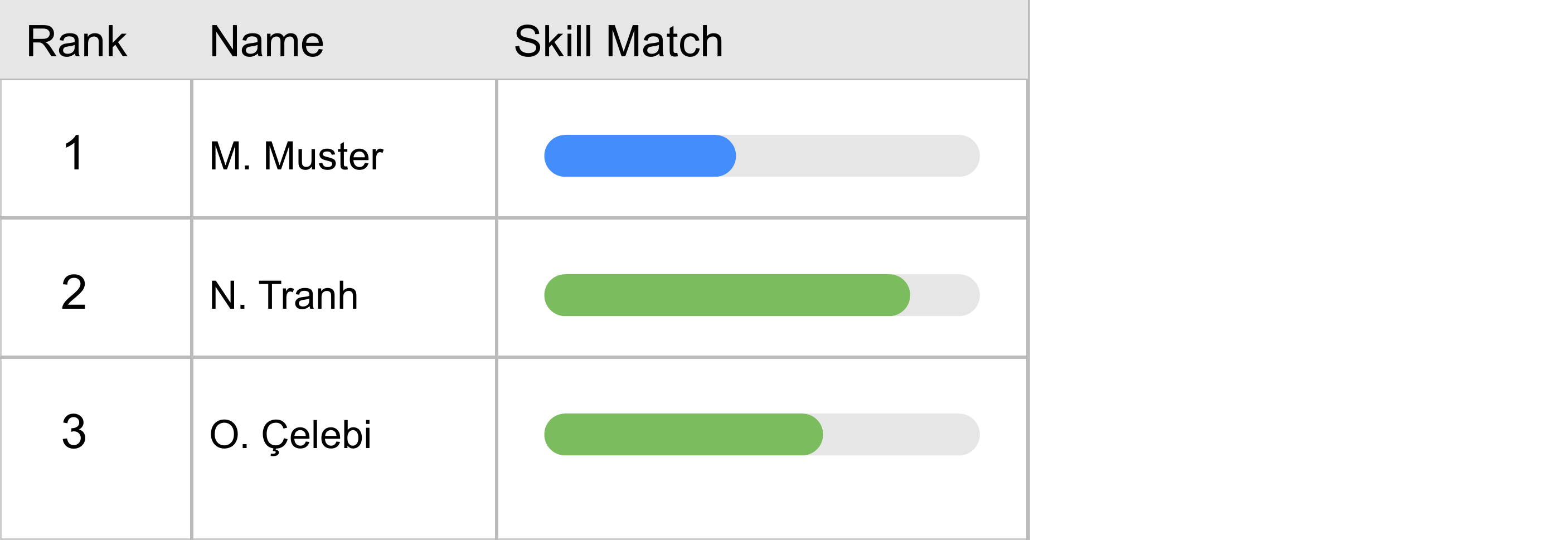}
		\caption{Baseline UI (BL)}
	\end{subfigure}
	\hfill
	\begin{subfigure}{0.45\textwidth}
		\includegraphics[width=\textwidth]{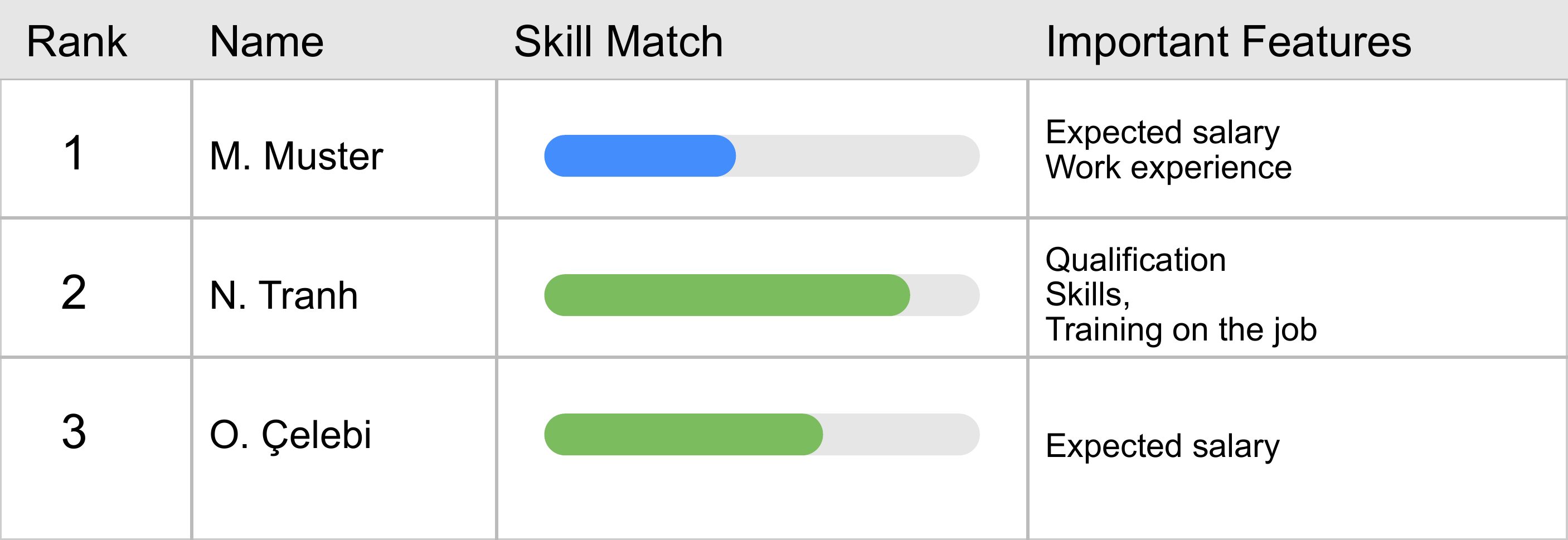}
		\caption{Feature importance UI}
	\end{subfigure}
	\begin{subfigure}{0.45\textwidth}
		\includegraphics[width=\textwidth]{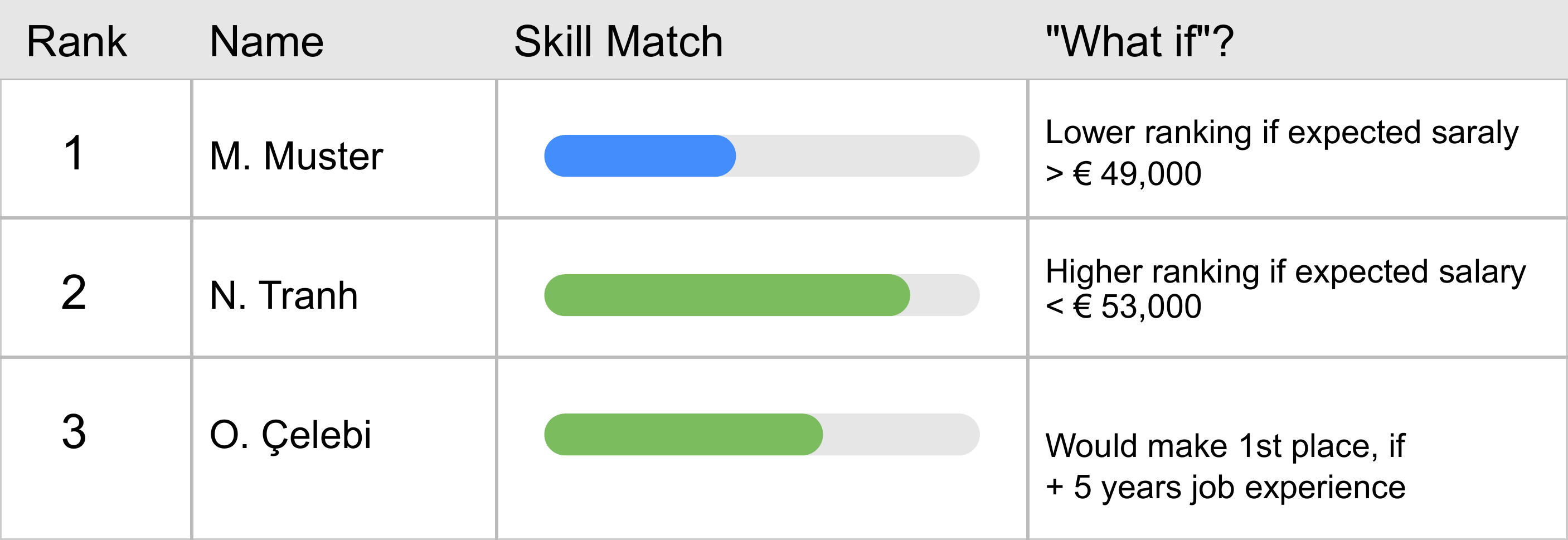}
		\caption{Counterfactuals UI}
	\end{subfigure}
	\hfill
	\begin{subfigure}{0.45\textwidth}
		\includegraphics[width=\textwidth]{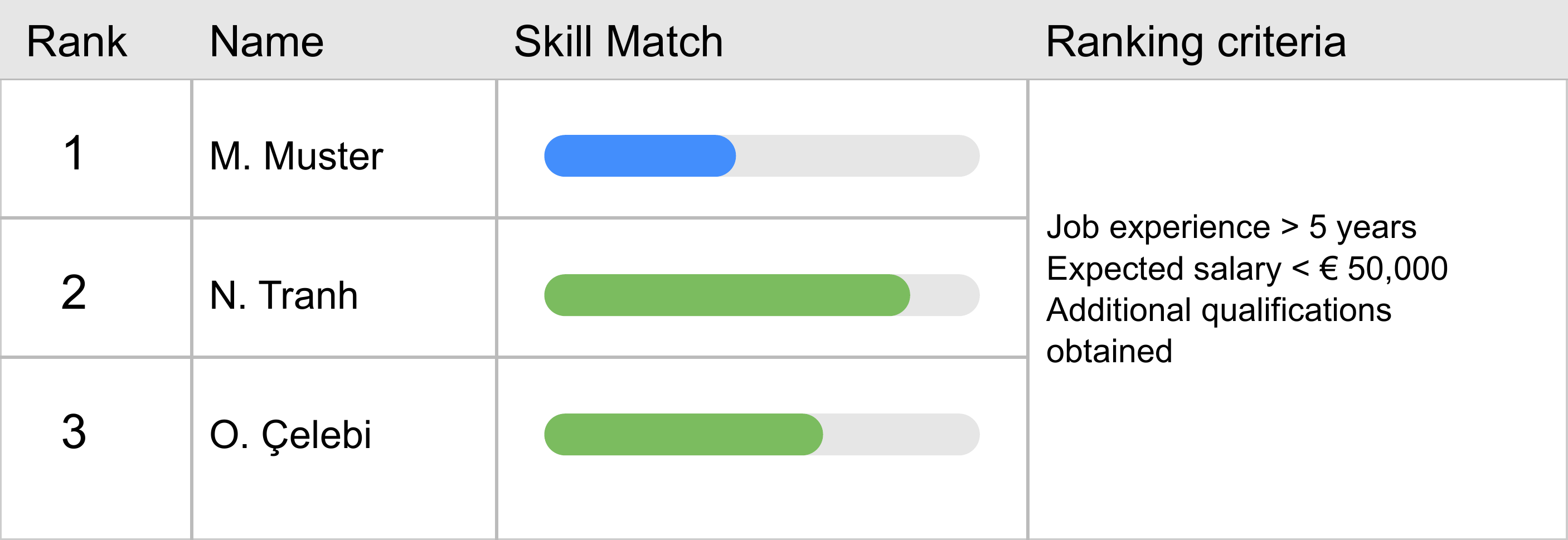}
		\caption{Model criteria UI}
	\end{subfigure}
	\caption{Recreated parts of the interface of Figure~\ref{fig:dashboards}. Without explanations (a), with feature importance (important features) (b), counterfactuals (what-if?) (c), and model criteria (ranking criteria) (d).}
	\label{fig:xai}
\end{figure*} 

\begin{table}[!bth] 
	\caption{Overview of the index AI Literacy} 
	\label{tab:ailit} 
	\begin{tabularx}{\columnwidth}{cccXc} 
		\toprule
		Scale & Items & Cronbach's $\alpha$ & Mean~(SD) & \textbf{$mean~(r_{it.corr})$}\\ 
		\midrule
		TU	&	5	&	$0.92$	&	$2.85$ ($1.28$)	&	$0.79$\\
		CA	&	5	&	$0.90$	&	$3.29$ ($1.22$)	&	$0.75$\\
		PA	&	5	&	$0.89$	&	$3.10$ ($1.24$)	&	$0.73$\\
		\bottomrule
	\end{tabularx} 
\end{table} 

\begin{table}[!bth]
	\caption{Overview of the AI Literacy groups} 
	\label{tab:ailitgroups} 
	\begin{tabularx}{\columnwidth}{Xccc} 
		\toprule
		AI Literacy & N & Mean (SD) & Median\\ 
		\midrule
		Low & $91$	&	$0.32$ ($0.09$) & $0.31$ \\
		Medium &	$185$	&	$0.61$ ($0.07$) & $0.60$\\
		High &	$134$ &	$0.83$ ($0.08$) &	$0.81$\\
		\bottomrule
	\end{tabularx} 
\end{table} 

We recreated the dashboard and experimental derivations using \textsc{Moqups} and added XAI elements to explain its AI results. Overall, the majority of (advertised as) AI dashboards contained no explicit information or warning about the results being AI-generated. Furthermore, the proposed metrics to assess, for example, performance, retention chances, or churn risks, lack further explanation. If the AI-based recommender systems in the tools used in practice resemble the illustrative materials, ambiguities are bound to occur. AI elements are utilized without further explanation of their core function, data sources, operations, or results, making the need for appropriate, targeted, and reliable XAI even more urgent.

The experiment focused on a ranking system that provides a recommendation (ranking) of incoming applications. From the advertisement material, the AI's ranking decision remains in need of explanation. Our experiment addressed this issue: we provided three different XAI-enhanced versions of the interface that each contained a different type of explanation---feature importance to assess the influencing factors on the results, counterfactuals to assess the decision boundaries of the model, and general model criteria to understand the meta-reasoning of the model (Figure~\ref{fig:xai}). To facilitate understanding among non-technical professionals, we referred to the XAI elements used in our experiments using more accessible terms: \enquote{Important features} (FI), \enquote{What if?} (CF), and \enquote{Ranking Criteria} (MC).

We measured \textit{subjective perception} using five Likert-scaled items to assess the perceived trustworthiness, transparency, comprehensibility, usefulness, and practical capability to act on the information provided by the participants. The items constituted a highly reliable indicator for subjective perception with consistently high Cronbach’s $\alpha$ values for the baseline dashboard and all XAI enhanced dashboards (cf. Table~\ref{tab:subjective}). 

\begin{table}[!h] 
	\caption{Overview of the index subjective perception} 
	\label{tab:subjective}
	\begin{tabularx}{\columnwidth}{lccXc} 
		\toprule
		& $N$  & Cronbach's $\alpha$ & Mean~(SD) & \textbf{$mean~(r_{it.corr})$}\\ 
		\midrule
		BL			&	410	&	$0.92$	&	$2.92$ ($1.21$)	&	$0.79$\\
		FI	&	136	&		$0.92$	&	$3.31$ ($1.11$)	&	$0.80$\\
		CF		&	138	&		$0.90$	&	$3.10$ ($1.14$)	&	$0.75$\\
		MC		&	136	&		$0.91$	&	$3.11$ ($1.10$)	&	$0.78$\\
		\bottomrule
	\end{tabularx} 
\end{table} 

\textit{Objective understanding} was operationalized as the number of correct responses to five factual statements derived from the information displayed on the dashboards (for example, “The person in second place has more suitable skills,” or “The data foundation is known.”). These questions could be answered with “yes,” “no,” or “you can't tell.” The last option indicated that the information presented on the dashboard did not support a definitive yes or no answer. This approach enabled us to award points for correct answers and thereby assess whether individuals could correctly interpret AI dashboards. Using this design, we examined how AI literacy influences objective understanding. By incorporating XAI elements, we evaluated their effectiveness by comparing the number of correct answers and drawing conclusions about XAI's impact across different levels of AI literacy.

We randomly assigned participants to three groups. All participants first evaluated the baseline user interface without explanations. Subsequently, each group was randomly assigned to assess a second interface with a specific explanation type. This randomization was implemented to prevent selection bias and systematic differences between groups.

\section{Findings}\label{sec:findings}
\subsection{Subjective perceptions of the interfaces}
\begin{figure*}
	\begin{subfigure}{0.48\textwidth}
		\includegraphics[width=\textwidth]{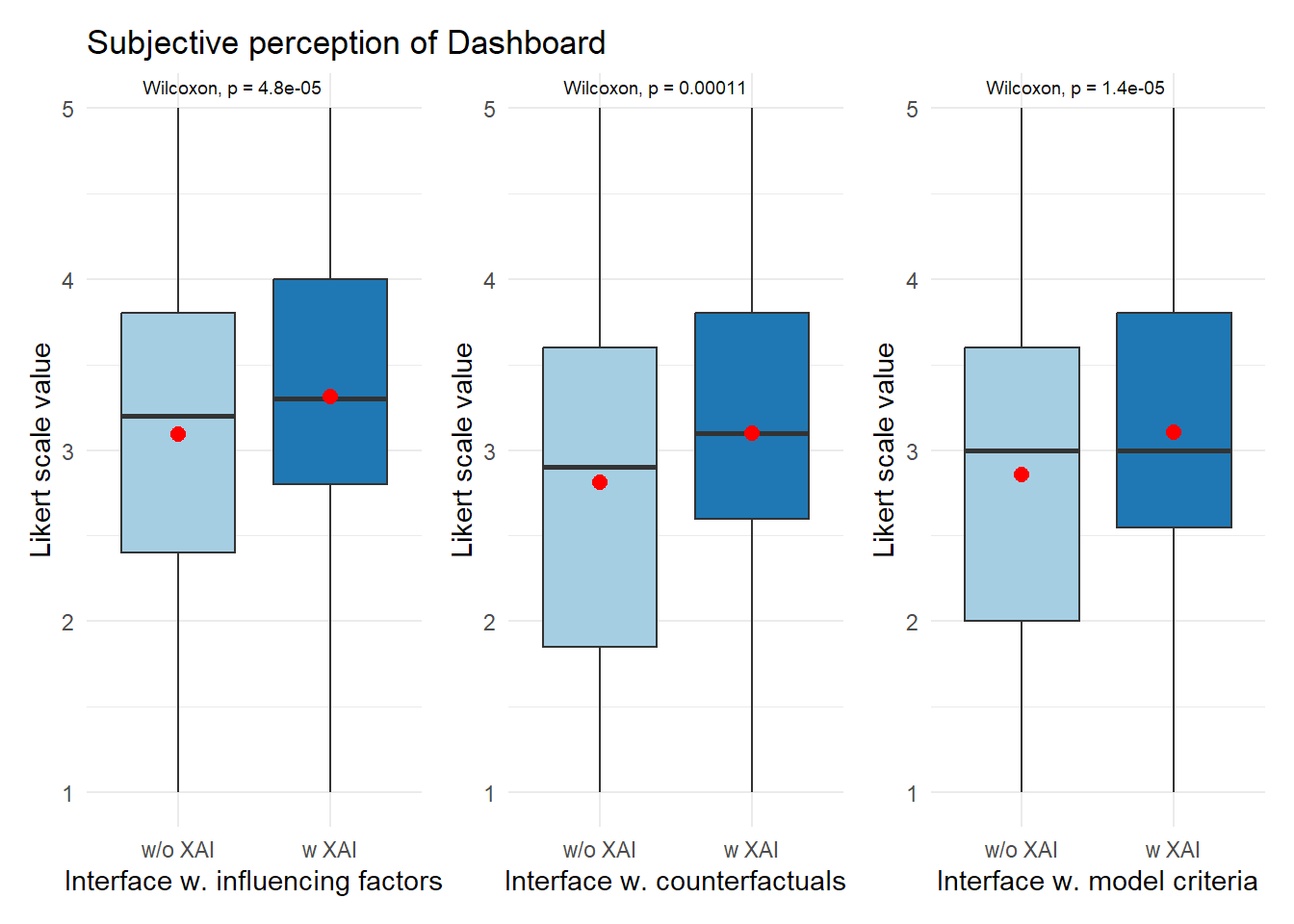}
		\caption{}
	\end{subfigure}
	\hfill
	\begin{subfigure}{0.48\textwidth}
		\includegraphics[width=\textwidth]{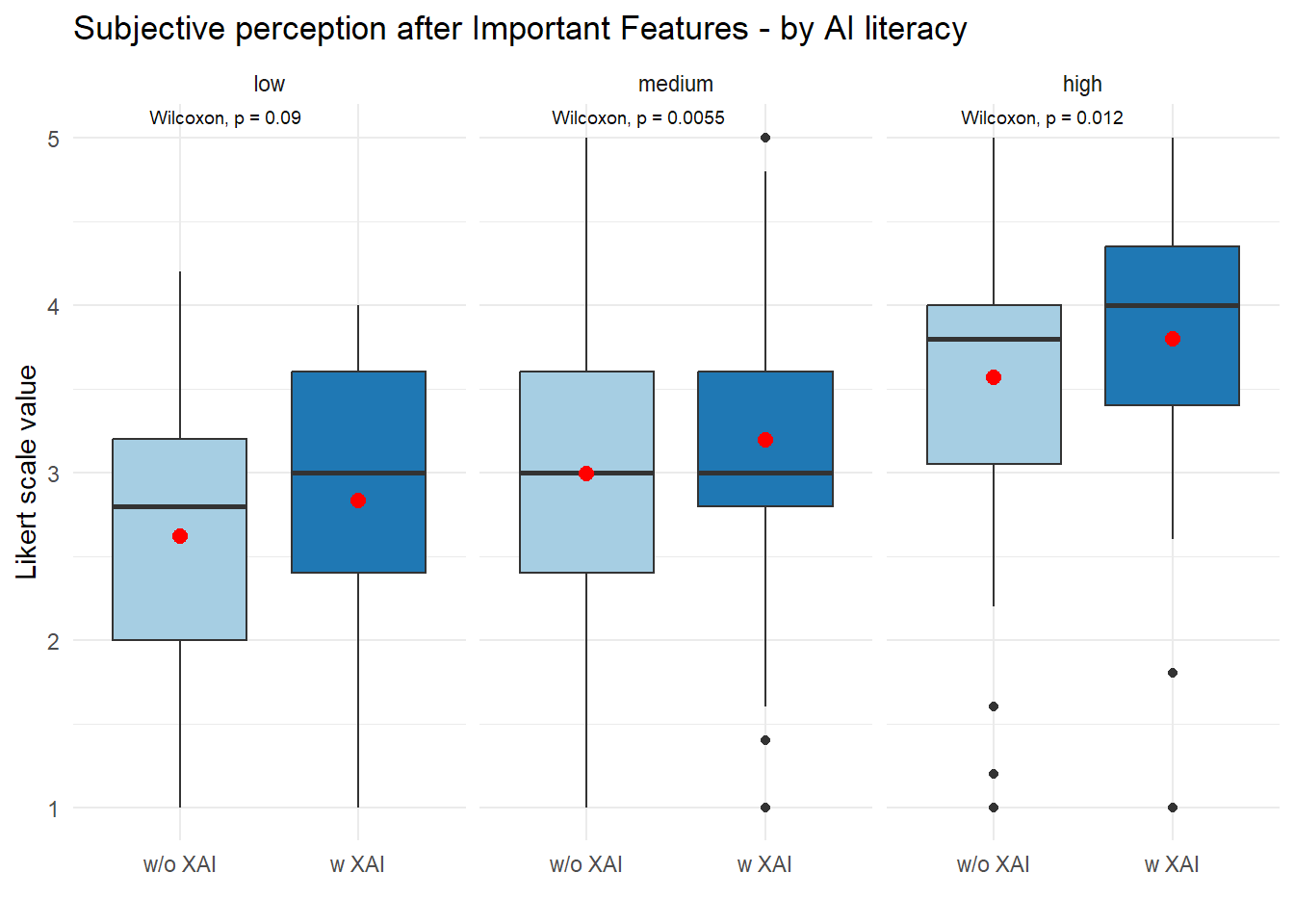}
		\caption{}
	\end{subfigure}
	\begin{subfigure}{0.48\textwidth}
		\includegraphics[width=\textwidth]{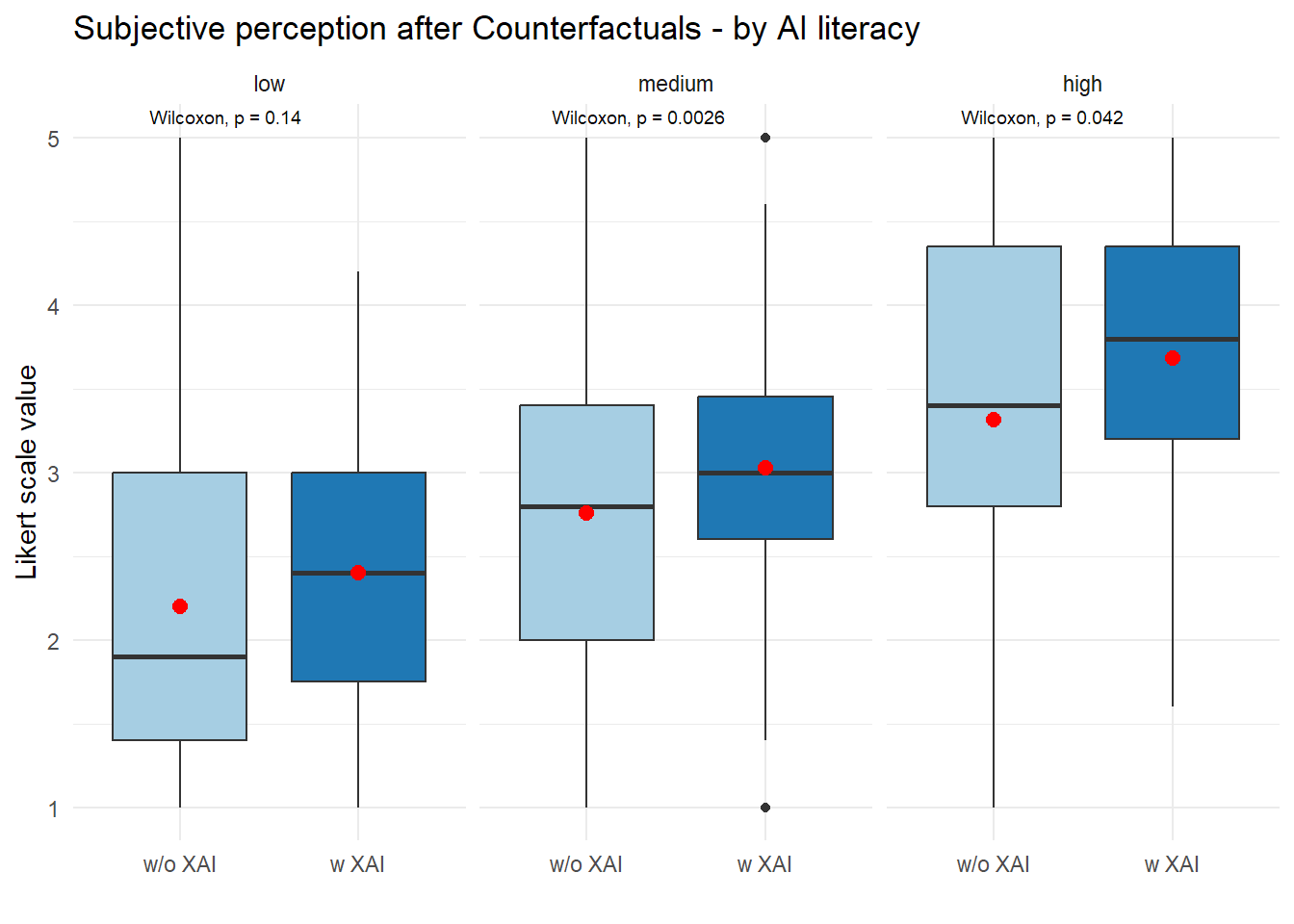}
		\caption{}
	\end{subfigure}
	\hfill
	\begin{subfigure}{0.48\textwidth}	
		\includegraphics[width=\textwidth]{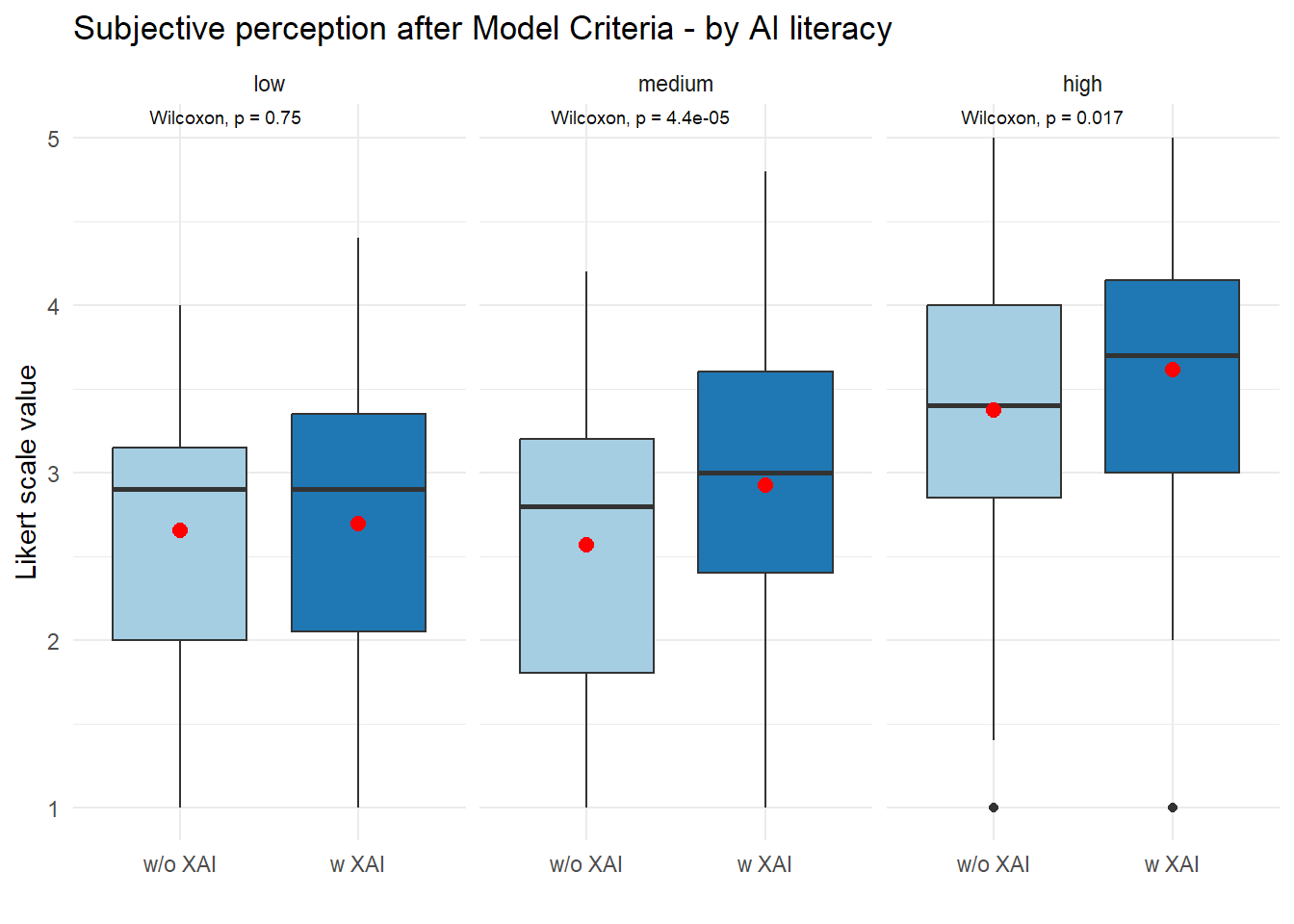}
		\caption{}
	\end{subfigure}
	\caption{Subjective perception of different XAI interfaces. Without differentiation by AI literacy (a). XAI elements differentiated by AI literacy (b-d). Red dot: $mean$ value.}
	\label{fig:subj}
\end{figure*}

The dashboards were first evaluated based on subjective perception by HR managers (Figure~\ref{fig:subj}a). To compare the subjective perception with and without XAI element for each literacy group, we conducted paired-sample Wilcoxon tests, because the results of the Likert scales for baseline and XAI-enhanced interfaces were non-normally distributed (Shapiro-Wilk test: Baseline $W = 0.972$, $p < .001$; XAI $W = 0.974$, $p < .001$). 

Adding any of the three XAI components to the interface yielded a consistent, though modest, upward shift in users’ subjective perception. Mean ratings for the \enquote{important features} interface rose from roughly $3.10$ without XAI to $3.31$ with XAI, for \enquote{counterfactuals} from $2.81$ to $3.10$, and for \enquote{model criteria} from $2.85$ to $3.11$. Paired‐samples Wilcoxon tests confirmed that all three increases were statistically robust ($p = 4.8 × 10^{–5}$, $p = 1.1 × 10^{–4}$, and $p = 1.4 × 10^{–5}$, respectively), indicating that the addition of explanatory information produced a small-to-moderate positive effect on perceived dashboard quality across the board. 

When we segmented participants by AI literacy (low, medium, high), however, the effect of XAI explanations was concentrated among users with at least moderate scores on the SNAIL scale (Figure~\ref{fig:subj}b-d). Low literacy users saw slight mean increases of $0.2–0.3$ points in all three interfaces, but none of these changes reached significance (influencing factors $p = .09$; counterfactuals $p = .14$; model criteria $p = .75$). Medium literacy users exhibited clear gains across every condition: the influencing‐factors interface increased by about $0.3$ points ($p = .0055$), counterfactuals by $0.4$ points ($p = .0026$), and model criteria by $0.5$ points ($p = 4.4 × 10^{–5}$). High-literacy users displayed a comparable pattern, with all three effects reaching significance (influencing factors, $p = .012$; counterfactuals, $p = .042$; model criteria, $p = .017$), and mean improvements of roughly $0.3–0.4$ points. 

Taken together, these results suggest that explanatory interfaces---in the form of influencing factors, counterfactuals, or explicit model criteria---systematically elevate users’ subjective perceptions of a user interface, if the recipient's AI literacy is at least moderate. Crucially, the benefit is most pronounced among individuals who already possess medium or high levels of AI literacy, whereas novices derive less measurable perceived usefulness. XAI elements do not compensate for low AI literacy and do not raise the group's subjective perception of the user interface's perceived quality. This suggests that users require prior knowledge to achieve subjective improvements in any explanation---and that explanation types for low AI literacy levels must be constructed differently.

\subsection{Objective understanding of the interfaces}
\begin{figure*}
	\begin{subfigure}{0.48\textwidth}
		\includegraphics[width=\textwidth]{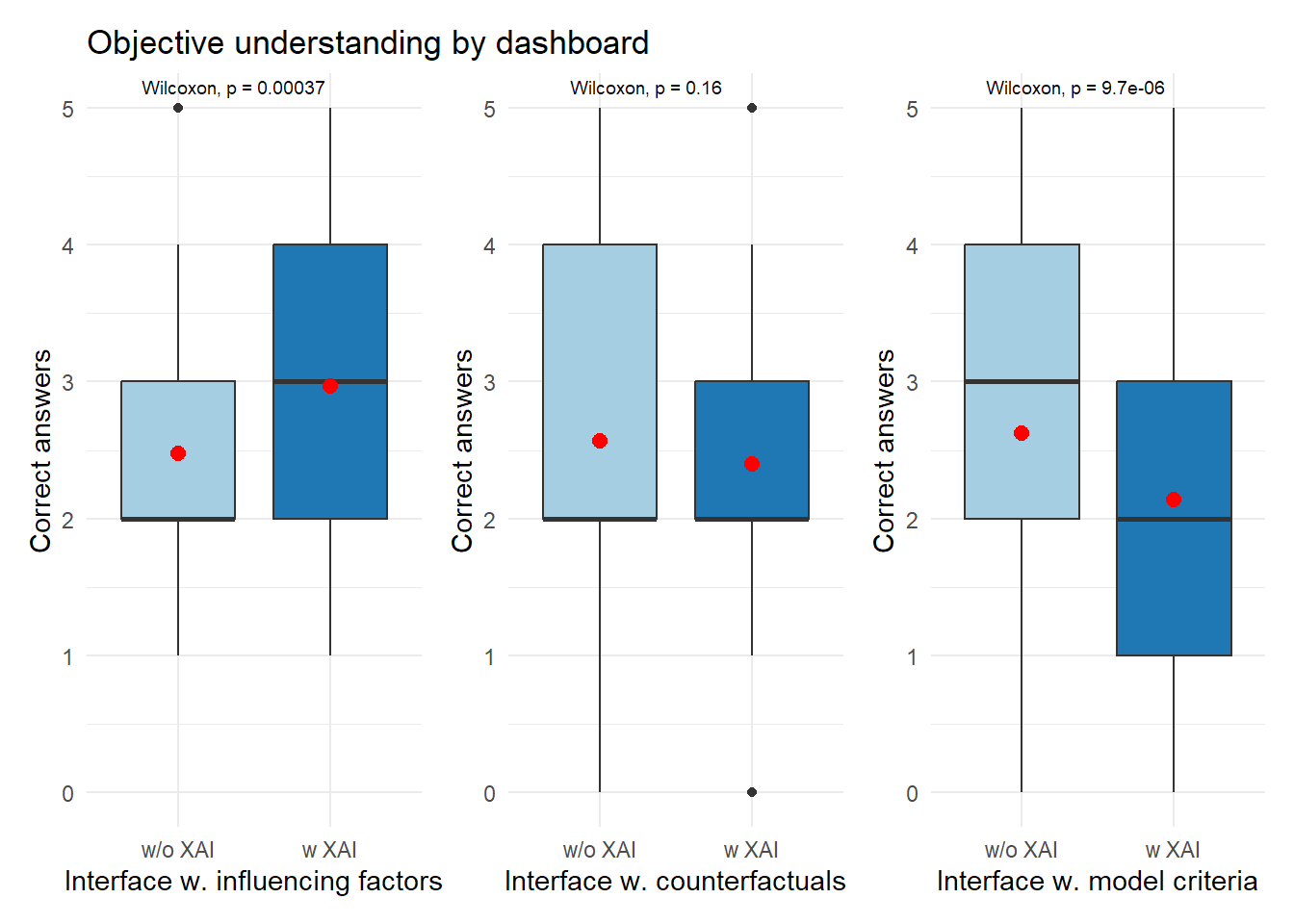}
		\caption{}
	\end{subfigure}
	\hfill
	\begin{subfigure}{0.48\textwidth}
		\includegraphics[width=\textwidth]{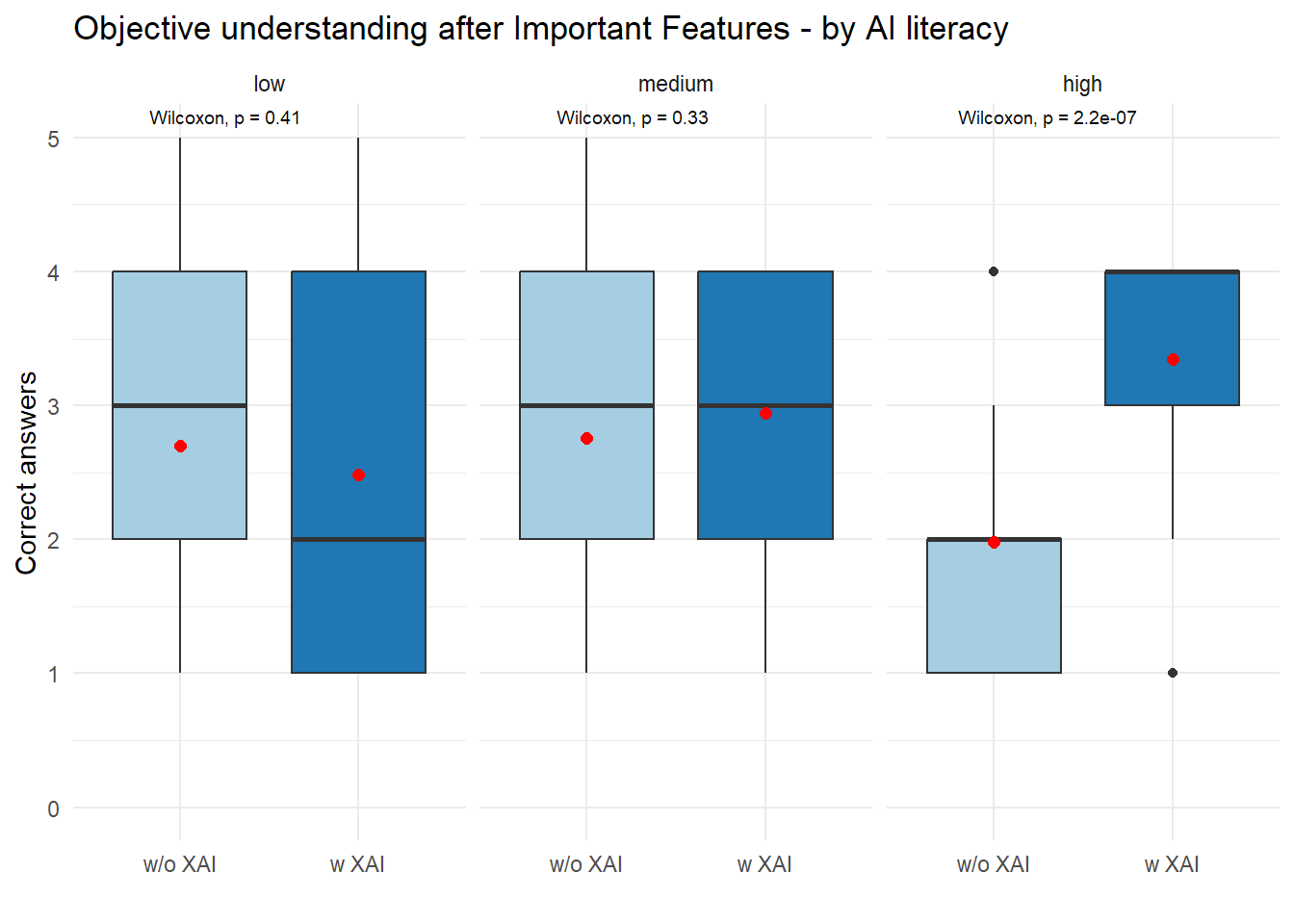}
		\caption{}
	\end{subfigure}
	\begin{subfigure}{0.48\textwidth}
		\includegraphics[width=\textwidth]{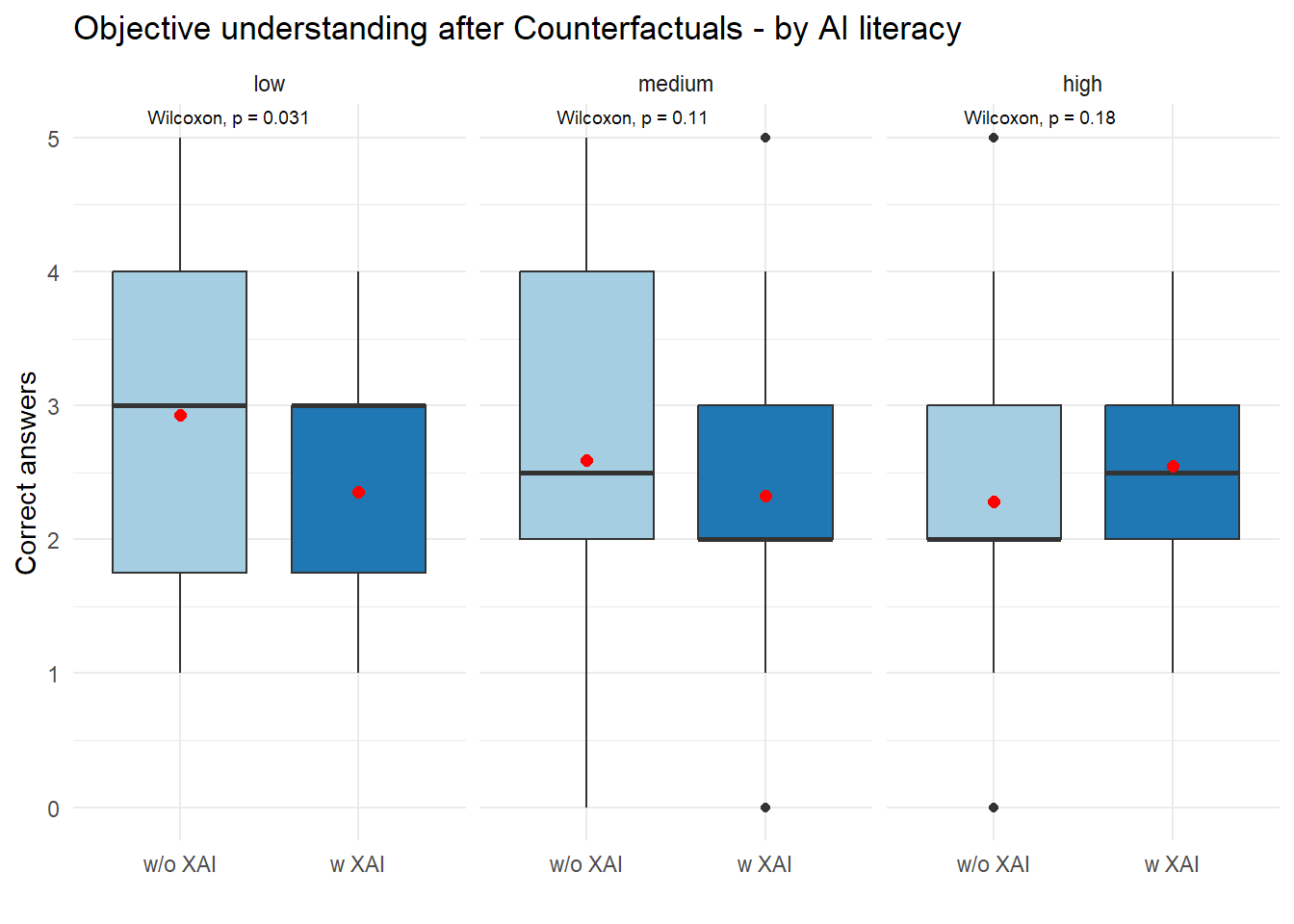}
		\caption{}
	\end{subfigure}
	\hfill
	\begin{subfigure}{0.48\textwidth}	
		\includegraphics[width=\textwidth]{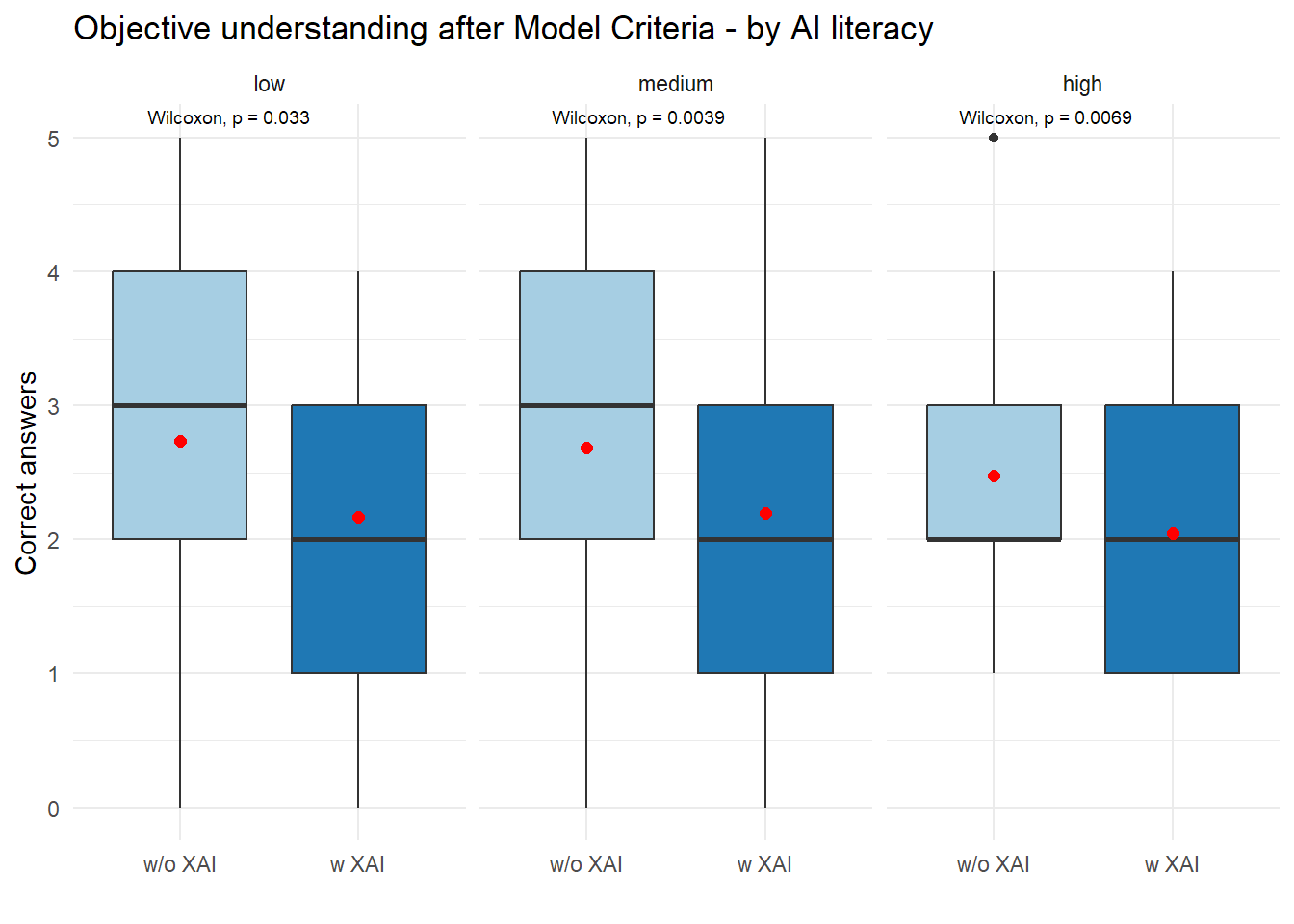}
		\caption{}
	\end{subfigure}
	\caption{Objective understanding of XAI interfaces. Without differentiation by AI literacy (a). XAI elements differentiated by AI literacy (b-d). Red dot: $mean$ value.}
	\label{fig:obj}
\end{figure*}

In the second part of the experiment, we examined how the three XAI elements influenced participants’ objective understanding of dashboard outputs. We again stratified these results by AI literacy (see Figure~\ref{fig:obj}). To compare performance with and without each XAI overlay within each literacy group, we conducted paired-sample Wilcoxon tests, because the scoring for baseline and each XAI-enhanced dashboard was non-normally distributed (Shapiro-Wilk test: Baseline $W = 0.910$, $p < .001$; XAI $W = 0.904$, $p < .001$). 

The XAI elements that show influencing factors on AI results overall performed best, since it was the only one that could increase the scores to the baseline dashboards. When participants were presented with enriched interfaces that displayed relevant features for the assessment scores, low‐ and medium‐literacy users exhibited negligible changes in their interpretation scores (low: mean $2.70$~vs.~$2.48$,~$p~=~.41$; medium: mean $2.75$ vs. $2.95$, $p = .33$), indicating that additionally represented features neither aided nor hindered their objective understanding. By contrast, high literacy users showed a significant improvement, with mean values rising from $1.98$ to $3.35$ ($p = 2.2 × 10^{-07}$). This large, highly significant effect suggests that only those with profound knowledge of AI were able to translate feature‐importance annotations into more accurate data interpretations.

The results are astonishing and indicate that the (randomized) group of high AI literacy might have been subjected to a systematic bias: the group performed worst in comparison to all other groups, especially to other randomized groups of high AI literacy, where equal scores would have been expected. The randomization of the experimental groups was successful with respect to the control variable AI literacy (Kruskal-Wallis test $p = .73$). However, significant differences were observed between the experimental groups in the baseline measurement of the dependent variables for subjective perception ($p = .03$), but non-significant differences for objective understanding ($p = .54$). Such discrepancies may arise despite randomization due to random variation.

Introducing counterfactuals as \enquote{if‐then} instances led to mixed outcomes. Low literacy participants performed worse when counterfactuals were present (mean $2.93$ vs. $2.36$, $p = .031$), a small but statistically significant decline. Medium literacy users showed a non-significant downward trend (mean $2.59$ vs. $2.32$, $p = .11$), and high literacy users remained essentially unchanged (mean $2.29$ vs. $2.55$, $p = .18$). These results suggest that novices may find counterfactual information distracting or confusing, while more experienced individuals neither consistently benefit nor suffer. 

Finally, overlaying explicit model criteria as information on what the model deems essential for the presented recommendations and decisions proved to be counterproductive across the board. Low literacy users’ interpretation scores fell from a mean of $2.73$ to $2.17$ ($p = .03$), medium literacy from $2.68$ to $2.20$ ($p = .004$), and high literacy from $2.48$ to $2.04$ ($p = .007$). All three decreases were statistically significant and of moderate effect size, indicating that detailed model criteria overwhelmed users regardless of their AI background, leading to poorer objective understanding.

Taken together, these findings demonstrate that XAI elements do not uniformly enhance users’ comprehension of dashboard information. While influencing factors can significantly boost accurate interpretations for technically savvy users, counterfactuals and explicit model criteria may impair or fail to improve objective understanding---particularly among those with limited AI literacy. 

\section{Discussion}\label{sec:discussion}
Our study provides novel empirical evidence on how AI literacy shapes HR managers’ interpretations of AI-based recruitment recommender systems. Addressing our first research question, we found that XAI elements can increase the quality of AI interfaces, depending on users’ AI literacy levels. However, enhanced subjective perceptions of informativeness, trustworthiness, and interpretability were statistically significant only for participants with at least moderate AI literacy. This suggests that users must possess foundational conceptual and critical skills to experience benefits from explainability elements in recruitment interfaces \cite{Laupichler.2023}. Low-literacy users gained little subjective value from XAI, in line with research that emphasizes the need for contextualized, domain-specific AI literacy interventions, and potentially confusing and misleading complexity \cite{Lintner.2024, Pinski.2024}.

Turning to our second research question, we observed a paradox: XAI explanations did not uniformly improve, and in some cases, impaired objective understanding. Only the \enquote{important features (feature importance)} element improved performance, but primarily for high literacy users. Counterfactual and model-criteria explanations either had no effect or reduced objective understanding, particularly among participants with low and medium literacy. Notably, high literacy users---while more likely to benefit subjectively from explanations---exhibited lower absolute understanding, which could be a sign of overconfidence in their AI literacy. This finding complicates earlier claims about the universal efficacy of XAI in fostering trust and better decisions \cite{Bhatt.2020, Speith.2022}, and supports concerns about information overload and cognitive miscalibration of non-technical users \cite{Bauer.2023}.

Our results align with and extend the literature on the intersection of AI literacy and XAI. The state of research has emphasized tailoring explanations to users’ backgrounds and roles \cite{Cambria.2023, Chowdhury.2023a}, but empirical investigations of explanation efficacy across literacy gradients in HR remain sparse. Our results indicate that designers should carefully tailor explanation formats to the target audience’s expertise level, avoiding overly complex explanations or too granular details. Our data suggest that XAI elements, when implemented in HR dashboards, may reinforce subjective confidence without reliably supporting accurate and responsible decision-making. This risk is amplified by regulatory demands for explainability and transparency in high-stakes contexts, such as those imposed by the EU AI Act and GDPR \cite{EuropeanCommission.2021, EuropeanParliament.2016}. These findings underline the importance of a nuanced approach to deploying AI tools in sensitive contexts, one that carefully balances explanation design with the end user’s AI knowledge. First, our baseline interface shows no clues to AI origins, data quality, or assumptions, leaving users to judge recommendations without deeper insights into underlying models or AI design decisions. Second, while XAI annotations enhance perceived information quality, they do not automatically translate into improved decision-making based on the recommender system. By contrasting subjective perceptions with objective understanding across literacy levels, we uncover the complex interplay between user expertise and explanation efficacy.

Upon further examination, we discovered that the type of explanation is crucial. Only additional important features---inspired by feature importance models without considering effect strengths---yielded an overall improvement in objective understanding performance. In contrast, counterfactual explanations and model‐criteria summaries hindered users’ evaluations. Counterfactuals---\enquote{What would have to change for the AI’s decision to differ?}---might impose substantial cognitive demands because they frame reasoning through a negated, hypothetical scenario rather than presenting a direct rationale. This result challenges prior claims about the universal effectiveness of counterfactuals \cite{MazzineBarbosadeOliveira.2023}. Likewise, model-criteria explanations---which merely translate the AI’s decision rules into human‐like justification---systematically failed, regardless of participants’ literacy.

Finally, our data reveal an overconfidence effect among high-literacy users: their strong self-assurance contradicts their performance in accurately showing objective understanding. This suggests that relying on self-reported AI literacy may obscure critical performance gaps; future research should incorporate objective measures of AI knowledge. This mirrors work in metacognition and digital literacy, suggesting that self-assessment may not be a reliable proxy for AI literacy, i.e., proper comprehension, critical capacity, or practical application \cite{Ng.2024, Lintner.2024}. For HRM practice, this implies that both AI literacy training and explanation design must be empirically validated for effectiveness.

We conclude that XAI is not a one-size-fits-all remedy in riskful settings like HR. Explanations must be carefully tailored to users’ expertise and meet the specific demands of their domain. However, providing several explanations at once to satisfy different AI literacy needs could raise the complexity of HR interfaces even further. Our study challenges the assumption that XAI elements are universally beneficial in HR recommender systems. Their impact is conditional on users’ AI literacy, the type of explanation, and the context of use. Effective, responsible deployment of XAI in HR, therefore, requires an integrated approach: robust, context-sensitive literacy training, careful user-centered explanation design, and ongoing evaluation of both subjective perception and objective understanding outcomes.

\section{Conclusion and Future Work}\label{sec:conclusion}
This research showed that HR managers’ AI literacy fundamentally affects the effectiveness of explainable AI elements in recruiting recommender systems. While XAI features enhance perceived transparency and trust among more literate users, they do not guarantee improved objective understanding, sometimes even undermining it. These findings call for a more nuanced, empirically grounded approach to the design and implementation of XAI in HRM. Specifically, we find that \textit{XAI is not a universal remedy} to make AI, its functions, and results accessible to non-technical professionals. Explanation strategies must be tailored to users’ actual literacy and cognitive needs. \textit{AI literacy is critical} and investments in AI for HR should be coupled with targeted, domain-specific literacy initiatives that inform and teach on specific tools and utilities and their mechanisms, like recommender systems. Finally, the design of explanations should \textit{address diverse interpretative needs} with flexible formats to serve users with heterogeneous backgrounds, while not overwhelming or misleading them, or giving them a false sense of understanding.

Future research should address several open questions: How can AI literacy interventions best be integrated into HR training programs, and what pedagogical approaches are most effective for non-technical professionals?
What hybrid or adaptive explanation strategies (for example, layered explanations, user-driven customization) can accommodate different literacy levels without increasing interface complexity?
How do explanation effects evolve with repeated exposure, feedback, or organizational learning?
What are the organizational and regulatory implications of overconfidence or miscalibration in AI-literate HR professionals? By pursuing these avenues, we can move toward HR recommender systems that are not only technically robust and legally compliant but also meaningfully transparent, fair, and supportive of human expertise. All in all, these results and the future outlook on open research emphasize that applied AI---in any domain---needs explanations. Explainability is a transdisciplinary process that involves technical interfaces, pedagogical and psychological learning capabilities, and social, vocational, or professional standards of specific groups, like HR managers.

\begin{acknowledgments}
	The research was part of the project \enquote{TRANKI -- Standards for transparent AI}, funded by the Hans Böckler Foundation (Grant no.: 2022-797-2). 
\end{acknowledgments}

\section*{Declaration on Generative AI}
During the preparation of this work, the author(s) used OpenAI's GPT-4.1 and o4-mini, as well as Grammarly, to check grammar and spelling and to enhance the writing style. After using these tools/services, the authors reviewed and edited the content as needed and take full responsibility for the publication’s content.

\bibliography{err}

\begin{thebibliography}{34}
\expandafter\ifx\csname natexlab\endcsname\relax\def\natexlab#1{#1}\fi
\providecommand{\url}[1]{\texttt{#1}}
\providecommand{\href}[2]{#2}
\providecommand{\path}[1]{#1}
\providecommand{\DOIprefix}{doi:}
\providecommand{\ArXivprefix}{arXiv:}
\providecommand{\URLprefix}{URL: }
\providecommand{\Pubmedprefix}{pmid:}
\providecommand{\doi}[1]{\href{http://dx.doi.org/#1}{\path{#1}}}
\providecommand{\Pubmed}[1]{\href{pmid:#1}{\path{#1}}}
\providecommand{\bibinfo}[2]{#2}
\ifx\xfnm\relax \def\xfnm[#1]{\unskip,\space#1}\fi
\bibitem[{Zhai et~al.(2024)Zhai, Zhang, and Yu}]{Zhai.2024}
\bibinfo{author}{Y.~Zhai}, \bibinfo{author}{L.~Zhang}, \bibinfo{author}{M.~Yu},
\newblock \bibinfo{title}{{{AI}} in {{Human Resource Management}}: {{Literature
  Review}} and {{Research Implications}}},
\newblock \bibinfo{journal}{Journal of the Knowledge Economy}
  (\bibinfo{year}{2024}). \DOIprefix\doi{10.1007/s13132-023-01631-z}.
\bibitem[{Malik et~al.(2023)Malik, Budhwar, and Kazmi}]{Malik.2023}
\bibinfo{author}{A.~Malik}, \bibinfo{author}{P.~Budhwar},
  \bibinfo{author}{B.~A. Kazmi},
\newblock \bibinfo{title}{Artificial intelligence ({{AI}})-assisted {{HRM}}:
  {{Towards}} an extended strategic framework},
\newblock \bibinfo{journal}{Human Resource Management Review}
  \bibinfo{volume}{33} (\bibinfo{year}{2023}) \bibinfo{pages}{100940}.
  \DOIprefix\doi{10.1016/j.hrmr.2022.100940}.
\bibitem[{Drage and Mackereth(2022)}]{Drage.2022}
\bibinfo{author}{E.~Drage}, \bibinfo{author}{K.~Mackereth},
\newblock \bibinfo{title}{Does {{AI Debias Recruitment}}? {{Race}}, {{Gender}},
  and {{AI}}'s "{{Eradication}} of {{Difference}}"},
\newblock \bibinfo{journal}{Philosophy \& Technology} \bibinfo{volume}{35}
  (\bibinfo{year}{2022}) \bibinfo{pages}{1--25}.
  \DOIprefix\doi{10.1007/s13347-022-00543-1}.
\bibitem[{Kelan(2024)}]{Kelan.2024}
\bibinfo{author}{E.~K. Kelan},
\newblock \bibinfo{title}{Algorithmic inclusion: {{Shaping}} the predictive
  algorithms of artificial intelligence in hiring},
\newblock \bibinfo{journal}{Human Resource Management Journal}
  \bibinfo{volume}{34} (\bibinfo{year}{2024}) \bibinfo{pages}{694--707}.
  \DOIprefix\doi{10.1111/1748-8583.12511}.
\bibitem[{Edwards et~al.(2024)Edwards, Charlwood, Guenole, and
  Marler}]{Edwards.2024}
\bibinfo{author}{M.~R. Edwards}, \bibinfo{author}{A.~Charlwood},
  \bibinfo{author}{N.~Guenole}, \bibinfo{author}{J.~Marler},
\newblock \bibinfo{title}{{{HR}} analytics: {{An}} emerging field finding its
  place in the world alongside simmering ethical challenges},
\newblock \bibinfo{journal}{Human Resource Management Journal}
  \bibinfo{volume}{34} (\bibinfo{year}{2024}) \bibinfo{pages}{326--336}.
  \DOIprefix\doi{10.1111/1748-8583.12435}.
\bibitem[{Kl{\"o}pper and K{\"o}hne(2023)}]{Klopper.2023}
\bibinfo{author}{M.~Kl{\"o}pper}, \bibinfo{author}{S.~K{\"o}hne},
\newblock \bibinfo{title}{Shifting {{Structures}} -- a systematic {{Literature
  Review}} on {{People Analytics}} and the {{Future}} of {{Work}}},
\newblock \bibinfo{journal}{ECIS 2023 Research Papers}  (\bibinfo{year}{2023})
  \bibinfo{pages}{1--19}.
\bibitem[{{European Commission}(2021)}]{EuropeanCommission.2021}
\bibinfo{author}{{European Commission}}, \bibinfo{title}{Regulation of the
  {{European Parliament}} and of the {{Council}} Laying down Harmonised
  {{Rules}} on {{Artificial Intelligence}} ({{Artificial Intelligence Act}})
  and Amending Certain {{Union}} Legislative {{Acts}}},
  \bibinfo{type}{Technical Report} \bibinfo{number}{2021/0106 (COD)},
  \bibinfo{address}{Brussels}, \bibinfo{year}{2021}.
\bibitem[{{European Parliament}(2016)}]{EuropeanParliament.2016}
\bibinfo{author}{{European Parliament}}, \bibinfo{title}{General {{Data
  Protection Regulation}}: {{GDPR}}}, \bibinfo{year}{2016}.
\bibitem[{Du(2024)}]{Du.2024}
\bibinfo{author}{J.~Du},
\newblock \bibinfo{title}{Exploring {{Gender Bias}} and {{Algorithm
  Transparency}}: {{Ethical Considerations}} of {{AI}} in {{HRM}}},
\newblock \bibinfo{journal}{Journal of Theory and Practice of Management
  Science} \bibinfo{volume}{4} (\bibinfo{year}{2024}) \bibinfo{pages}{36--43}.
  \DOIprefix\doi{10.53469/jtpms.2024.04(03).06}.
\bibitem[{Fabris et~al.(2024)Fabris, Baranowska, Dennis, Graus, Hacker,
  Saldivar, Zuiderveen~Borgesius, and Biega}]{Fabris.2024}
\bibinfo{author}{A.~Fabris}, \bibinfo{author}{N.~Baranowska},
  \bibinfo{author}{M.~J. Dennis}, \bibinfo{author}{D.~Graus},
  \bibinfo{author}{P.~Hacker}, \bibinfo{author}{J.~Saldivar},
  \bibinfo{author}{F.~Zuiderveen~Borgesius}, \bibinfo{author}{A.~J. Biega},
\newblock \bibinfo{title}{Fairness and {{Bias}} in {{Algorithmic Hiring}}: {{A
  Multidisciplinary Survey}}},
\newblock \bibinfo{journal}{ACM Transactions on Intelligent Systems and
  Technology}  (\bibinfo{year}{2024}). \DOIprefix\doi{10.1145/3696457}.
\bibitem[{Simbeck(2019)}]{Simbeck.2019}
\bibinfo{author}{K.~Simbeck},
\newblock \bibinfo{title}{{{HR}} analytics and {{Ethics}}},
\newblock \bibinfo{journal}{IBM Journal of Research and Development}
  \bibinfo{volume}{63} (\bibinfo{year}{2019}) \bibinfo{pages}{9:1--9:12}.
  \DOIprefix\doi{10.1147/JRD.2019.2915067}.
\bibitem[{K{\"o}chling et~al.(2021)K{\"o}chling, Riazy, Wehner, and
  Simbeck}]{Kochling.2021}
\bibinfo{author}{A.~K{\"o}chling}, \bibinfo{author}{S.~Riazy},
  \bibinfo{author}{M.~C. Wehner}, \bibinfo{author}{K.~Simbeck},
\newblock \bibinfo{title}{Highly {{Accurate}}, {{But Still Discriminatory}}:
  {{A Fairness Evaluation}} of {{Algorithmic Video Analysis}} in the
  {{Recruitment Context}}},
\newblock \bibinfo{journal}{Business \& Information Systems Engineering}
  \bibinfo{volume}{63} (\bibinfo{year}{2021}) \bibinfo{pages}{39--54}.
  \DOIprefix\doi{10.1007/s12599-020-00673-w}.
\bibitem[{Molnar(2022)}]{Molnar.2022}
\bibinfo{author}{C.~Molnar}, \bibinfo{title}{Interpretable Machine Learning:
  {{A}} Guide for Making Black Box Models Explainable}, \bibinfo{edition}{2}
  ed., \bibinfo{publisher}{Selfpublishing}, \bibinfo{address}{Munich},
  \bibinfo{year}{2022}.
\bibitem[{Bhatt et~al.(2020)Bhatt, Xiang, Sharma, Weller, Taly, Jia, Ghosh,
  Puri, Moura, and Eckersley}]{Bhatt.2020}
\bibinfo{author}{U.~Bhatt}, \bibinfo{author}{A.~Xiang},
  \bibinfo{author}{S.~Sharma}, \bibinfo{author}{A.~Weller},
  \bibinfo{author}{A.~Taly}, \bibinfo{author}{Y.~Jia},
  \bibinfo{author}{J.~Ghosh}, \bibinfo{author}{R.~Puri},
  \bibinfo{author}{J.~M.~F. Moura}, \bibinfo{author}{P.~Eckersley},
\newblock \bibinfo{title}{Explainable machine learning in deployment},
\newblock in: \bibinfo{editor}{M.~Hildebrandt} (Ed.),
  \bibinfo{booktitle}{Proceedings of the 2020 {{Conference}} on {{Fairness}},
  {{Accountability}}, and {{Transparency}}}, \bibinfo{publisher}{Association
  for Computing Machinery}, \bibinfo{address}{New York}, \bibinfo{year}{2020},
  pp. \bibinfo{pages}{648--657}. \DOIprefix\doi{10.1145/3351095.3375624}.
\bibitem[{Speith(2022)}]{Speith.2022}
\bibinfo{author}{T.~Speith},
\newblock \bibinfo{title}{A {{Review}} of {{Taxonomies}} of {{Explainable
  Artificial Intelligence}} ({{XAI}}) {{Methods}}},
\newblock in: \bibinfo{editor}{{Association for Computing Machinery}} (Ed.),
  \bibinfo{booktitle}{{{FAccT}} '22: {{Proceedings}} of the 2022 {{ACM
  Conference}} on {{Fairness}}, {{Accountability}}, and {{Transparency}}},
  \bibinfo{publisher}{Association for Computing Machinery},
  \bibinfo{address}{New York}, \bibinfo{year}{2022}, pp.
  \bibinfo{pages}{2239--2250}. \DOIprefix\doi{10.1145/3531146.3534639}.
\bibitem[{Khalili(2023)}]{Khalili.2023}
\bibinfo{author}{M.~Khalili},
\newblock \bibinfo{title}{Against the opacity, and for a qualitative
  understanding, of artificially intelligent technologies},
\newblock \bibinfo{journal}{AI and Ethics}  (\bibinfo{year}{2023}).
  \DOIprefix\doi{10.1007/s43681-023-00332-2}.
\bibitem[{Bauer et~al.(2023)Bauer, Zahn, and Hinz}]{Bauer.2023}
\bibinfo{author}{K.~Bauer}, \bibinfo{author}{M.~Zahn},
  \bibinfo{author}{O.~Hinz},
\newblock \bibinfo{title}{Expl({{AI}})ned: {{The Impact}} of {{Explainable
  Artificial Intelligence}} on {{Users}}' {{Information Processing}}},
\newblock \bibinfo{journal}{Information Systems Research}
  (\bibinfo{year}{2023}). \DOIprefix\doi{10.1287/isre.2023.1199}.
\bibitem[{Laupichler et~al.(2023)Laupichler, Aster, Haverkamp, and
  Raupach}]{Laupichler.2023}
\bibinfo{author}{M.~C. Laupichler}, \bibinfo{author}{A.~Aster},
  \bibinfo{author}{N.~Haverkamp}, \bibinfo{author}{T.~Raupach},
\newblock \bibinfo{title}{Development of the ``{{Scale}} for the assessment of
  non-experts' {{AI}} literacy'' -- {{An}} exploratory factor analysis},
\newblock \bibinfo{journal}{Computers in Human Behavior Reports}
  \bibinfo{volume}{12} (\bibinfo{year}{2023}) \bibinfo{pages}{100338}.
  \DOIprefix\doi{10.1016/j.chbr.2023.100338}.
\bibitem[{Carolus et~al.(2023)Carolus, Koch, Straka, Latoschik, and
  Wienrich}]{Carolus.2023}
\bibinfo{author}{A.~Carolus}, \bibinfo{author}{M.~J. Koch},
  \bibinfo{author}{S.~Straka}, \bibinfo{author}{M.~E. Latoschik},
  \bibinfo{author}{C.~Wienrich},
\newblock \bibinfo{title}{{{MAILS}} - {{Meta AI}} literacy scale:
  {{Development}} and testing of an {{AI}} literacy {{Questionnaire}} based on
  well-founded {{Competency Models}} and psychological {{Change-}} and
  {{Meta-Competencies}}},
\newblock \bibinfo{journal}{Computers in Human Behavior: Artificial Humans}
  \bibinfo{volume}{1} (\bibinfo{year}{2023}) \bibinfo{pages}{100014}.
  \DOIprefix\doi{10.1016/j.chbah.2023.100014}.
\bibitem[{Guidotti et~al.(2019)Guidotti, Monreale, Ruggieri, Turini, Giannotti,
  and Pedreschi}]{Guidotti.2019}
\bibinfo{author}{R.~Guidotti}, \bibinfo{author}{A.~Monreale},
  \bibinfo{author}{S.~Ruggieri}, \bibinfo{author}{F.~Turini},
  \bibinfo{author}{F.~Giannotti}, \bibinfo{author}{D.~Pedreschi},
\newblock \bibinfo{title}{A {{Survey}} of {{Methods}} for {{Explaining Black
  Box Models}}},
\newblock \bibinfo{journal}{ACM Computing Surveys} \bibinfo{volume}{51}
  (\bibinfo{year}{2019}) \bibinfo{pages}{1--42}.
  \DOIprefix\doi{10.1145/3236009}.
\bibitem[{Bodria et~al.(2023)Bodria, Giannotti, Guidotti, Naretto, Pedreschi,
  and Rinzivillo}]{Bodria.2023}
\bibinfo{author}{F.~Bodria}, \bibinfo{author}{F.~Giannotti},
  \bibinfo{author}{R.~Guidotti}, \bibinfo{author}{F.~Naretto},
  \bibinfo{author}{D.~Pedreschi}, \bibinfo{author}{S.~Rinzivillo},
\newblock \bibinfo{title}{Benchmarking and survey of explanation methods for
  black box models},
\newblock \bibinfo{journal}{Data Mining and Knowledge Discovery}
  \bibinfo{volume}{37} (\bibinfo{year}{2023}) \bibinfo{pages}{1719--1778}.
  \DOIprefix\doi{10.1007/s10618-023-00933-9}.
\bibitem[{Bhat and Long(2024)}]{Bhat.2024}
\bibinfo{author}{M.~Bhat}, \bibinfo{author}{D.~Long},
\newblock \bibinfo{title}{Designing {{Interactive Explainable AI Tools}} for
  {{Algorithmic Literacy}} and {{Transparency}}},
\newblock in: \bibinfo{booktitle}{Designing {{Interactive Systems
  Conference}}}, \bibinfo{publisher}{ACM}, \bibinfo{address}{Copenhagen
  Denmark}, \bibinfo{year}{2024}, pp. \bibinfo{pages}{939--957}.
  \DOIprefix\doi{10.1145/3643834.3660722}.
\bibitem[{Lintner(2024)}]{Lintner.2024}
\bibinfo{author}{T.~Lintner},
\newblock \bibinfo{title}{A systematic review of {{AI}} literacy scales},
\newblock \bibinfo{journal}{NPJ science of learning} \bibinfo{volume}{9}
  (\bibinfo{year}{2024}) \bibinfo{pages}{50}.
  \DOIprefix\doi{10.1038/s41539-024-00264-4}.
\bibitem[{Ng et~al.(2024)Ng, Wu, Leung, Chiu, and Chu}]{Ng.2024}
\bibinfo{author}{D.~T.~K. Ng}, \bibinfo{author}{W.~Wu},
  \bibinfo{author}{J.~K.~L. Leung}, \bibinfo{author}{T.~K.~F. Chiu},
  \bibinfo{author}{S.~K.~W. Chu},
\newblock \bibinfo{title}{Design and validation of the {{AI}} literacy
  questionnaire: {{The}} affective, behavioural, cognitive and ethical
  approach},
\newblock \bibinfo{journal}{British Journal of Educational Technology}
  \bibinfo{volume}{55} (\bibinfo{year}{2024}) \bibinfo{pages}{1082--1104}.
  \DOIprefix\doi{10.1111/bjet.13411}.
\bibitem[{Pinski et~al.(2024)Pinski, Hofmann, and Benlian}]{Pinski.2024}
\bibinfo{author}{M.~Pinski}, \bibinfo{author}{T.~Hofmann},
  \bibinfo{author}{A.~Benlian},
\newblock \bibinfo{title}{{{AI Literacy}} for the top management: {{An}} upper
  echelons perspective on corporate {{AI}} orientation and implementation
  ability},
\newblock \bibinfo{journal}{Electronic Markets} \bibinfo{volume}{34}
  (\bibinfo{year}{2024}) \bibinfo{pages}{24}.
  \DOIprefix\doi{10.1007/s12525-024-00707-1}.
\bibitem[{Bassellier et~al.(2003)Bassellier, Benbasat, and
  Reich}]{Bassellier.2003}
\bibinfo{author}{G.~Bassellier}, \bibinfo{author}{I.~Benbasat},
  \bibinfo{author}{B.~H. Reich},
\newblock \bibinfo{title}{The {{Influence}} of {{Business Managers}}' {{IT
  Competence}} on {{Championing IT}}},
\newblock \bibinfo{journal}{Information Systems Research} \bibinfo{volume}{14}
  (\bibinfo{year}{2003}) \bibinfo{pages}{317--336}.
  \DOIprefix\doi{10.1287/isre.14.4.317.24899}.
\bibitem[{Clinciu and Hastie(2019)}]{Clinciu.2019}
\bibinfo{author}{M.-A. Clinciu}, \bibinfo{author}{H.~Hastie},
\newblock \bibinfo{title}{A {{Survey}} of {{Explainable AI Terminology}}},
\newblock \bibinfo{journal}{Proceedings of the 1st Workshop on Interactive
  Natural Language Technology for Explainable Artificial Intelligence (NL4XAI
  2019)}  (\bibinfo{year}{2019}) \bibinfo{pages}{8--13}.
  \DOIprefix\doi{10.18653/v1/W19-8403}.
\bibitem[{Gunning et~al.(2019)Gunning, Stefik, Choi, Miller, Stumpf, and
  Yang}]{Gunning.2019}
\bibinfo{author}{D.~Gunning}, \bibinfo{author}{M.~Stefik},
  \bibinfo{author}{J.~Choi}, \bibinfo{author}{T.~Miller},
  \bibinfo{author}{S.~Stumpf}, \bibinfo{author}{G.-Z. Yang},
\newblock \bibinfo{title}{{{XAI-Explainable}} artificial intelligence},
\newblock \bibinfo{journal}{Science robotics} \bibinfo{volume}{4}
  (\bibinfo{year}{2019}). \DOIprefix\doi{10.1126/scirobotics.aay7120}.
\bibitem[{Chowdhury et~al.(2023)Chowdhury, {Joel-Edgar}, Dey, Bhattacharya, and
  Kharlamov}]{Chowdhury.2023a}
\bibinfo{author}{S.~Chowdhury}, \bibinfo{author}{S.~{Joel-Edgar}},
  \bibinfo{author}{P.~K. Dey}, \bibinfo{author}{S.~Bhattacharya},
  \bibinfo{author}{A.~Kharlamov},
\newblock \bibinfo{title}{Embedding transparency in artificial intelligence
  machine learning models: Managerial implications on predicting and explaining
  employee turnover},
\newblock \bibinfo{journal}{The International Journal of Human Resource
  Management} \bibinfo{volume}{34} (\bibinfo{year}{2023})
  \bibinfo{pages}{2732--2764}. \DOIprefix\doi{10.1080/09585192.2022.2066981}.
\bibitem[{Cambria et~al.(2023)Cambria, Malandri, Mercorio, Mezzanzanica, and
  Nobani}]{Cambria.2023}
\bibinfo{author}{E.~Cambria}, \bibinfo{author}{L.~Malandri},
  \bibinfo{author}{F.~Mercorio}, \bibinfo{author}{M.~Mezzanzanica},
  \bibinfo{author}{N.~Nobani},
\newblock \bibinfo{title}{A survey on {{XAI}} and natural language
  explanations},
\newblock \bibinfo{journal}{Information Processing \& Management}
  \bibinfo{volume}{60} (\bibinfo{year}{2023}) \bibinfo{pages}{103111}.
  \DOIprefix\doi{10.1016/j.ipm.2022.103111}.
\bibitem[{Holzinger et~al.(2022)Holzinger, Saranti, Molnar, Biecek, and
  Samek}]{Holzinger.2022b}
\bibinfo{author}{A.~Holzinger}, \bibinfo{author}{A.~Saranti},
  \bibinfo{author}{C.~Molnar}, \bibinfo{author}{P.~Biecek},
  \bibinfo{author}{W.~Samek},
\newblock \bibinfo{title}{Explainable {{AI Methods}} - {{A Brief Overview}}},
\newblock in: \bibinfo{editor}{A.~Holzinger}, \bibinfo{editor}{R.~Goebel},
  \bibinfo{editor}{R.~Fong}, \bibinfo{editor}{T.~Moon}, \bibinfo{editor}{K.-R.
  M{\"u}ller}, \bibinfo{editor}{W.~Samek} (Eds.), \bibinfo{booktitle}{{{xxAI}}
  - {{Beyond Explainable AI}}: {{International Workshop}}, Held in Conjunction
  with {{ICML}} 2020}, \bibinfo{publisher}{Springer}, \bibinfo{address}{Cham},
  \bibinfo{year}{2022}, pp. \bibinfo{pages}{13--38}.
  \DOIprefix\doi{10.1007/978-3-031-04083-2_2}.
\bibitem[{K{\"u}hl et~al.(2023)K{\"u}hl, Meske, Nitsche, and
  Lobana}]{Kuhl.2023a}
\bibinfo{author}{N.~K{\"u}hl}, \bibinfo{author}{C.~Meske},
  \bibinfo{author}{M.~Nitsche}, \bibinfo{author}{J.~Lobana},
\newblock \bibinfo{title}{Investigating the {{Role}} of {{Explainability}} and
  {{AI Literacy}} in {{User Compliance}}},
\newblock \bibinfo{journal}{SSRN Electronic Journal}  (\bibinfo{year}{2023}).
  \DOIprefix\doi{10.2139/ssrn.4558966}.
\bibitem[{{Mazzine Barbosa de Oliveira} et~al.(2023){Mazzine Barbosa de
  Oliveira}, Goethals, Brughmans, and Martens}]{MazzineBarbosadeOliveira.2023}
\bibinfo{author}{R.~{Mazzine Barbosa de Oliveira}},
  \bibinfo{author}{S.~Goethals}, \bibinfo{author}{D.~Brughmans},
  \bibinfo{author}{D.~Martens}, \bibinfo{title}{Unveiling the {{Potential}} of
  {{Counterfactuals Explanations}} in {{Employability}}},
  \bibinfo{type}{Technical Report}, arXiv, \bibinfo{year}{2023}.
  \DOIprefix\doi{10.48550/arXiv.2305.10069}.
\bibitem[{Laupichler et~al.(2023)Laupichler, Aster, Perschewski, and
  Schleiss}]{Laupichler.2023a}
\bibinfo{author}{M.~C. Laupichler}, \bibinfo{author}{A.~Aster},
  \bibinfo{author}{J.-O. Perschewski}, \bibinfo{author}{J.~Schleiss},
\newblock \bibinfo{title}{Evaluating {{AI Courses}}: {{A Valid}} and {{Reliable
  Instrument}} for {{Assessing Artificial-Intelligence Learning}} through
  {{Comparative Self-Assessment}}},
\newblock \bibinfo{journal}{Education Sciences} \bibinfo{volume}{13}
  (\bibinfo{year}{2023}) \bibinfo{pages}{978}.
  \DOIprefix\doi{10.3390/educsci13100978}.

\end{thebibliography}

\appendix
\onecolumn
\section{Additional tables of indices statistics}
\begin{table}[!h] \centering 
	\caption{Selected items from \cite{Laupichler.2023, Laupichler.2023a} and individual item statistics} 
	\label{tab:items} 
	\begin{tabularx}{\textwidth}{cXcccc} 
		\toprule
		Item & Statement: \textit{I can \dots}	&	Mean (SD)	&	Item total $r$	& $\alpha$ if dropped\\ 
		\midrule
		TU 1 & explain how AI applications make decisions. & $2.92$ ($1.28$) & $0.80$ & $0.90$ \\ 
		TU 2 & explain the difference between general (or strong) and narrow (or weak) artificial intelligence. & $2.72$ ($1.31$) & $0.80$ & $0.90$ \\ 
		TU 3 & explain how machine learning works at a general level. & $2.85$ ($1.29$) & $0.81$ & $0.90$ \\ 
		TU 4 & describe the concept of explainable AI. & $2.88$ ($1.29$) & $0.79$ & $0.90$ \\ 
		TU 5 & evaluate whether media representations of AI (for example, in movies or video games) go beyond the current capabilities of AI technologies. & $2.89$ ($1.22$) & $0.76$ & $0.91$ \\ 
		\cmidrule(lr){1-5}
		CA 1 & explain why data privacy must be considered when developing and using artificial intelligence applications. & $3.46$ ($1.23$) & $0.72$ & $0.88$ \\ 
		CA 2 & identify ethical issues surrounding artificial intelligence. & $3.28$ ($1.19$) & $0.74$ & $0.88$ \\ 
		CA 3 & name weaknesses of artificial intelligence. & $3.23$ ($1.20$) & $0.74$ & $0.88$ \\ 
		CA 4 & describe potential legal problems that may arise when using artificial intelligence. & $3.16$ ($1.25$) & $0.75$ & $0.87$ \\ 
		CA 5 & explain why data plays an important role in the development and application of artificial intelligence. & $3.33$ ($1.22$) & $0.78$ & $0.87$ \\ 
		\cmidrule(lr){1-5}
		PA 1 & give examples from my daily life (personal or professional) where I might be in contact with artificial intelligence. & $3.31$ ($1.23$) & $0.71$ & $0.87$ \\ 
		PA 2 & tell if the technologies I use are supported by artificial intelligence. & $3.04$ ($1.24$) & $0.75$ & $0.86$ \\ 
		PA 3 & assess if a problem in my field can and should be solved with artificial intelligence methods. & $3.10$ ($1.22$) & $0.76$ & $0.86$ \\ 
		PA 4 & name applications in which AI-assisted natural language processing/understanding is used. & $2.80$ ($1.31$) & $0.72$ & $0.87$ \\ 
		PA 5 & critically evaluate the implications of artificial intelligence applications in at least one subject area. & $3.25$ ($1.21$) & $0.72$ & $0.87$ \\ 
		\bottomrule
	\end{tabularx} 
\end{table} 

\section{Screenshots of AI Dashboards}
\begin{figure}[h!]
	\includegraphics[width=\textwidth]{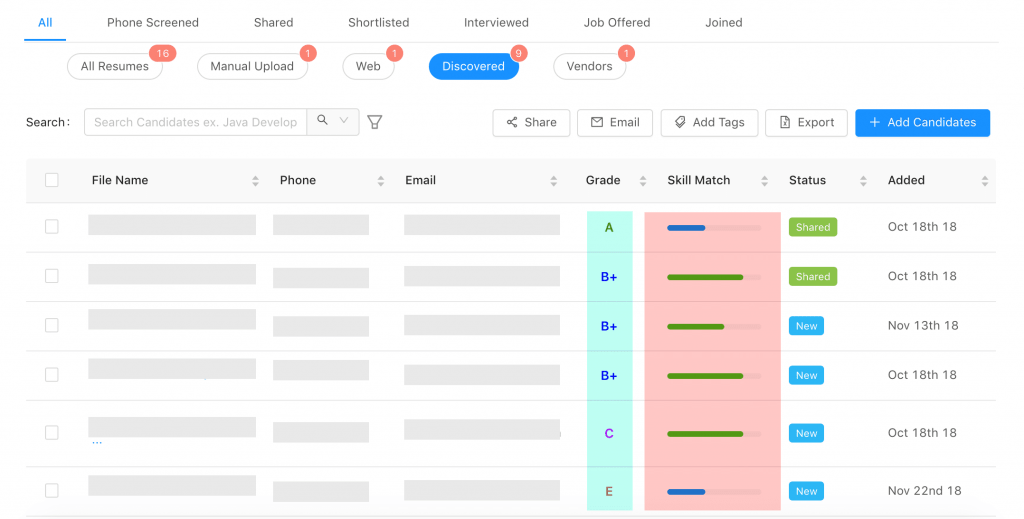}
	\caption{Resume screening and recommender system. Source: \url{https://cvviz.com/product/resume-screening/}.}
	\label{fig:dashboards}
\end{figure}

\begin{figure}
	\includegraphics[width=.95\textwidth]{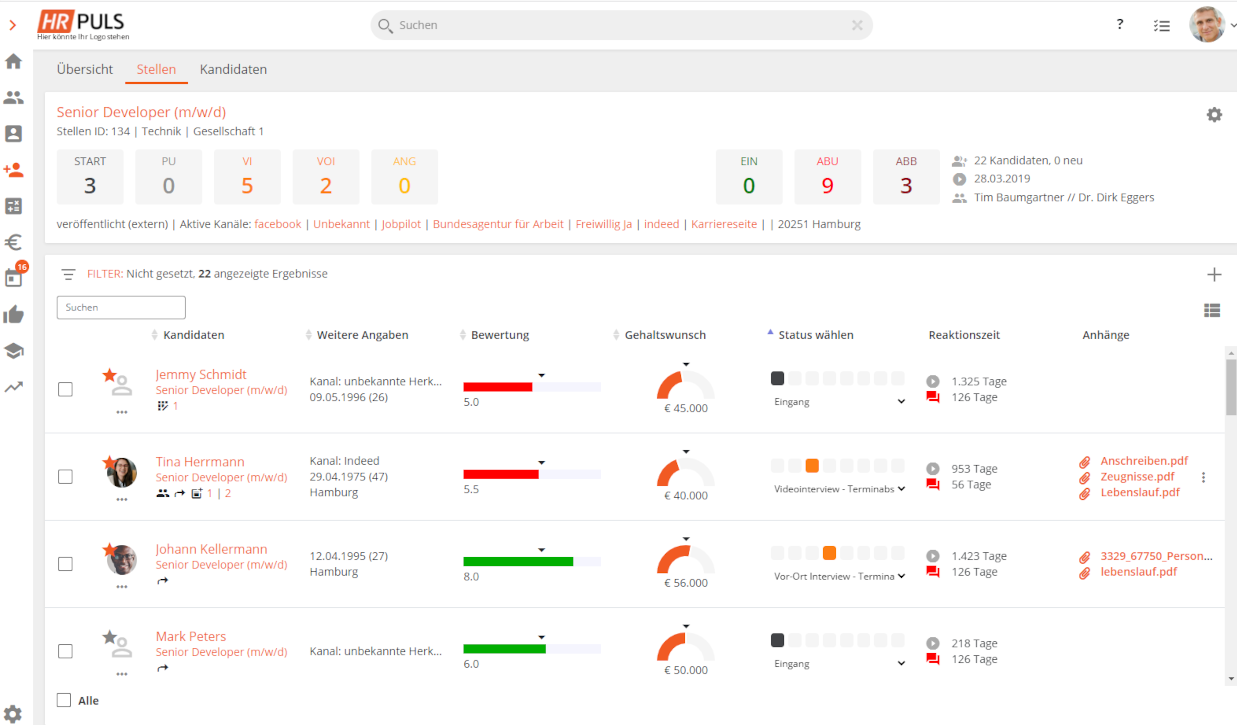}
	\caption{Resume screening and recommender system. Source: \url{https://www.hrpuls.de/bewerbermanagement.html}.}
	\label{fig:dashboards2}
\end{figure}

\begin{figure}
	\includegraphics[width=.95\textwidth]{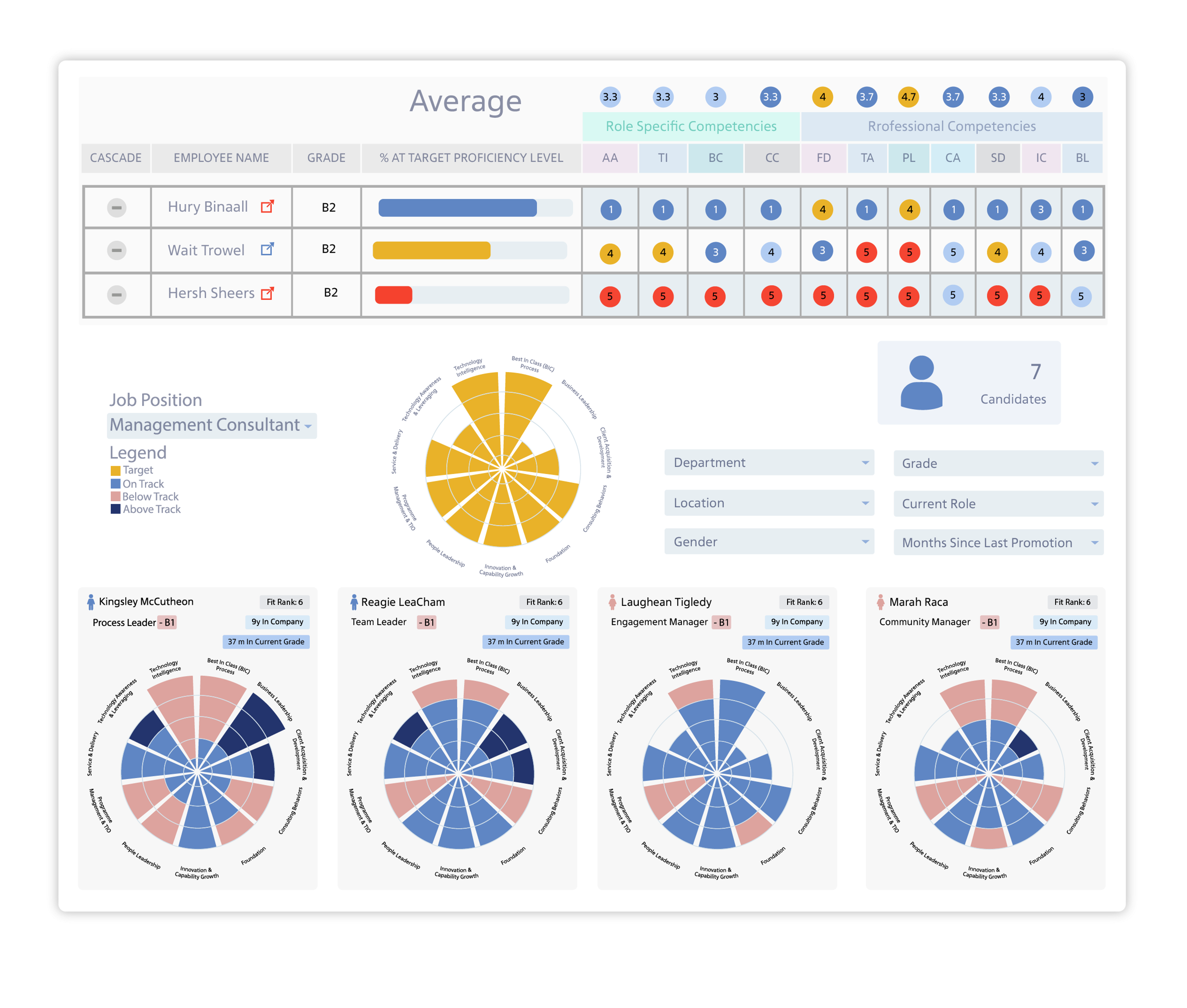}
	\caption{Talent management and development recommender system. Source: \url{https://www.softwareadvice.com/bi/edligo-talent-analytics-profile/}.}
	\label{fig:dashboards3}
\end{figure}
\end{document}